\documentclass[preprint,3p,12pt]{elsarticle}

\usepackage{graphicx, appendix}
\usepackage[colorlinks=true]{hyperref}
\usepackage{amssymb, bm, amsmath}
\usepackage{wrapfig}
\usepackage{float}
\usepackage{multirow}
\usepackage{array}
\usepackage{booktabs}
\usepackage{xcolor}
\usepackage{lipsum}
\usepackage{placeins}
\usepackage{float}
\biboptions{numbers,sort&compress}

\journal{Journal of the Mechanics and Physics of Solids}

\begin{document}

\begin{frontmatter}



\title{Bending Mechanics of Biomimetic Scale  Plates}


\author[inst1]{Pranta Rahman Sarkar}
\author[inst1]{Hossein Ebrahimi}
\author[inst1]{Md Shahjahan Hossain}
\author[inst1]{Ranajay Ghosh\corref{cor1}\corref{cor2}}
\ead{ranajay.ghosh@ucf.edu}

\cortext[cor1]{Corresponding author address: 4000 Central Florida Blvd, Orlando, Florida 32816, USA}
\cortext[cor2]{Corresponding author telephone: +1 407-823-3402}

\affiliation[inst1]{organization={Department of Mechanical and Aerospace Engineering, University of Central Florida},%Department and Organization
            %addressline={Address One}, 
            city={Orlando},
            %postcode={00000}, 
            state={FL},
            country={USA}}

\begin{abstract}
We develop the fundamentals of nonlinear and anisotropic bending behavior of biomimetic scale plates using a combination of analytical modeling, finite element (FE) computations, and motivational experiments. The analytical architecture-property relationships are derived for both synclastic and anticlastic curvatures. The results show that, as the scales engage, both synclastic and anticlastic deformations show non-linear scale contact kinematics and cross-curvature sensitivity of moments resulting in strong curvature-dependent elastic nonlinearity and emergent anisotropy. The anisotropy of bending rigidities and their evolution with curvature are affected by both the direction and magnitude of bending as well as scale geometry parameters, and their distribution on the substrate. Like earlier beam-like substrates, kinematic locked states were found to occur; however, their existence and evolution are also strongly determined by scale geometry and imposed cross-curvatures. This validated model helps us to quantify bending response, locking behavior, and their geometric dependence, paving the way for a deeper understanding of the nature of nonlinearity and anisotropy of these systems.

\end{abstract}


% \begin{highlights}
% \end{highlights}

\begin{keyword}
biomimetic scale \sep anisotropic plate  \sep nonlinear elasticity \sep tailorable response
\end{keyword}

\end{frontmatter}

% %%%%%%%%%%%%%%%%%%%%%%%%%%%%%%%%%%%%%%%%%%%%%%%%%%
% \vspace{-5pt}
% \begin{figure} [htbp]
% \begin{center}
% \includegraphics[scale = 0.5]{6-Fig1.pdf}
% \end{center}
% \vspace{-15pt}
% \caption
%  {(sample) (a) Natural fish scales under deformation mode. (b) Fabricated biomimetic scale metamaterials under bending deformation. (c) Schematic diagram of a biomimetic scale-covered plate subjected to bending load and RVE geometry.}
% \label{6-Fig1} 
% \end{figure}
% %%%%%%%%%%%%%%%%%%%%%%%%%%%%%%%%%%%%%%%%%%%%%%%%%%%
%%check test
% \begin{figure}[htbp]
%  \centering
%  \begin{tabular}{ccc}
%   (a) & (b) & (c) \\
%  \includegraphics[scale = 0.25]{Fish_scale copy.png} & \includegraphics[scale = 0.2]{Sample_side_v2.png} \\
%  %\includegraphics[scale = 0.25]{Figure_1_b.pdf}\\
%   (d) & (e) & (f) \\ \\ \includegraphics[scale = 0.048]{monoclastic.png} &
% \includegraphics[scale = 0.048]{synclastic_v3.png} & \includegraphics[scale = 0.048]{anticlastic_v2.png}
% \end{tabular}
%  %\vspace{-10pt}
% \caption{a) Natural Fish with Visible Scales
% b) Artificially Fabricated 2D Fish Scale Structure
% c) Schematic Illustration of Fish Dimensions
% d) Monoclastic Bending Phenomenon
% e) Synclastic Bending Effect
% f) Anticlastic Bending Behavior.}
%  \label{6-Fig1}
% \end{figure}
%%%%%%%%%%%%%%%%%%%%%%%%%%%%%%%%%%%%%%%%%%%%%%%%%%%

%%%%%%%%%%%%%%%%%%%%%%%%%%%%%%%%%%%%%%%
\begin{figure} [t]
\begin{center}
\includegraphics[scale = 0.48]{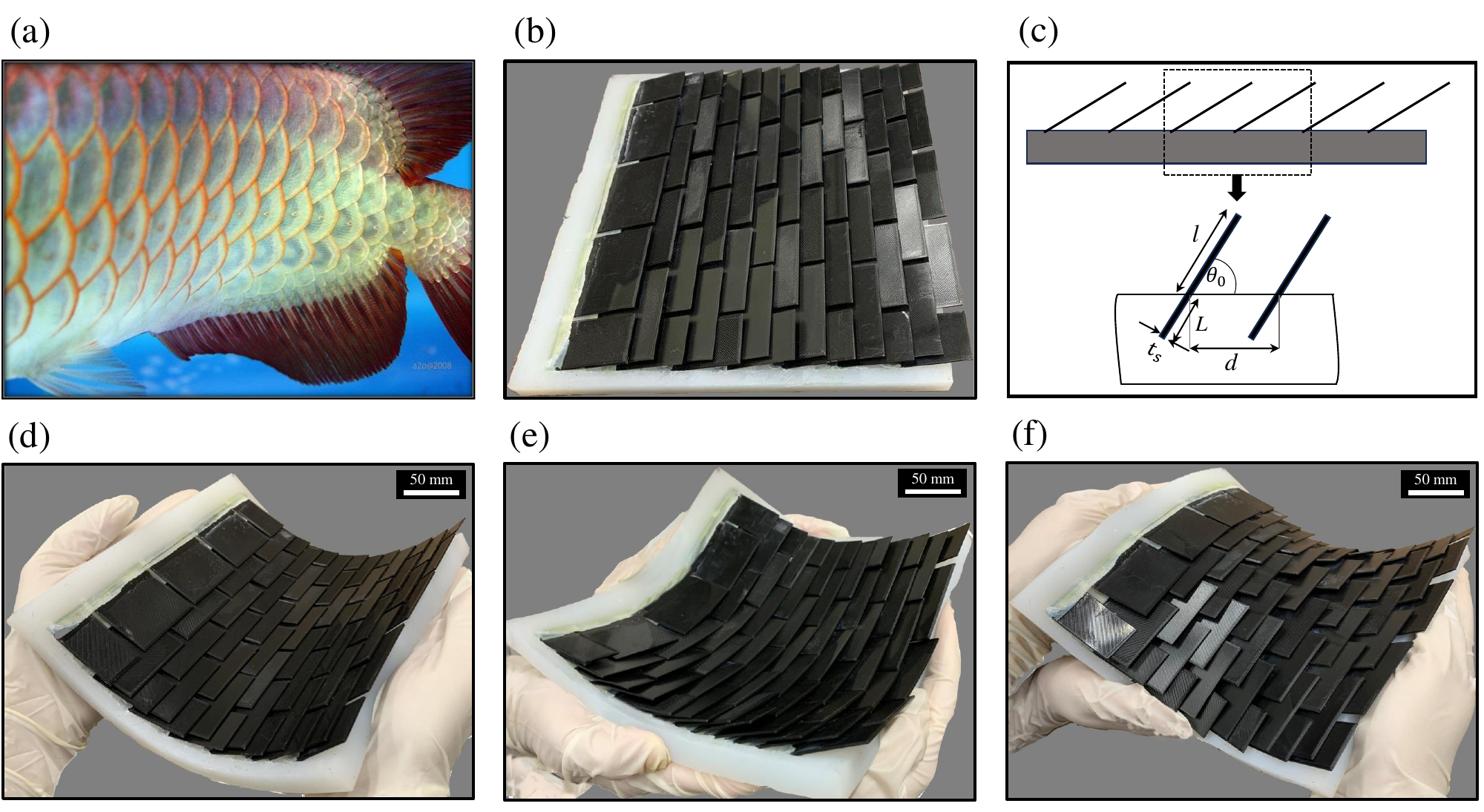}
\end{center}
{\caption {(a) Natural fish with visible scales. (b) Artificially fabricated 3D fish scale structure. (c) Schematic illustration of scale dimensions. (d) Monoclastic bending deformation of the scale-covered plate. (e) Synclastic bending deformation of the scale-covered plate. (f) Anticlastic bending deformation of the scale-covered plate.}}
\label{6-Fig1}
\end{figure}
%%%%%%%%%%%%%%%%%%%%%%%%%%%%%%%%%%%%%%%

\section{Introduction}

In nature, dermal scales such as those found on fish, reptile or even mammals can serve multiple functions including assistance in swimming, locomotion, camouflage, thermal regulation, and protection from predatory attacks  \cite{long1996functions,huang2006controlled, song2010quantitative, chen2011armadillo,chen2012predation, yang2013natural,sun2015fish, naleway2016structure}. Among these, it is now well understood that the mechanical advantages provided by fish scales are not merely from the additional ‘coating’ or 'thin film' effect of overlapping elements but reliant on nonlinear kinematics of scales sliding and their interplay with the deformation of the underlying substrate. Thus, their arrangement, engagement, overlap, orientations, and distributions become key elements that govern overall performance \cite{zhu2013intricate,song2011threat,zhu2012structure}. Early theoretical models have indicated that these arrangements lead to concerted sliding in periodic or locally periodic fashion amplifying the overall non-linearity causing significant strain stiffening that leads up to a quasi-rigid locked state \cite{vernerey2010mechanics,ghosh2014contact, ebrahimi2019tailorable}. Subsequent studies also indicate that these contact nonlinearities contribute to a number of other complex mechanical effects such as nonlinear elasticity of twisting \cite{ebrahimi2019tailorable}, puncture resistance \cite{martini2017comparative, browning2013mechanics}, fracture resistance \cite{murcia2015temperature}, dual nature of friction \cite{ghosh2016frictional}, and emergent viscosity in damping \cite{ebrahimi2023material}. Such behavior, distinct from traditional composites, can be further exploited to create tunable behavior substrates \cite{tatari2023bending}, buckling resistant structures \cite{shafiei2021bioinspired}, etc.

A substantial amount of early literature has been dedicated to isolating the influence of nonlinear contact kinematics in the genesis of nonlinear elasticity for 1D structures such as scale-covered beams. In most of these studies, the scales were typically kept rigid or allowed to bend only slightly (stiff scales). It was found that bending \cite{ghosh2014contact} and twisting \cite{ebrahimi2019tailorable} of a 1D scale-covered substrate lead to reversible nonlinear strain stiffening and interlocking (jamming) behavior even in small strains. Additionally, previous studies on the coupling of bending and twisting \cite{dharmavaram2022coupled} demonstrated how such cross-curvature coupling effects work together in fish scales, distinct from the behavior observed under individual loading conditions. This distinction is due to the complex engagement patterns between the scales that are often non-commutative. Although these effects do not need friction to arise, friction certainly amplifies their effect without fundamentally changing the contact-kinematics \cite{ghosh2016frictional,ebrahimi2020coulomb}. The overall broad contours of engagement were not changed even when periodicity conditions were relaxed \cite{ali2019bending,ali2019tailorable} or other non-ideal effects were incorporated \cite{ghosh2017non}, although the models gave better agreements with numerical simulations. The analytical models are considered indispensable for these systems due to severe shortcomings of the FE-based commercial computational codes. Recently, discrete element method (DEM) based computational analysis of scaled covered 1D substrate have been reported \cite{shafiei2021very,shafiei2021bioinspired} to overcome this impasse. \cite{karuriya2024plastic} implemented DEM to computationally analyze granular crystals which also shows excellent tunable mechanical properties. The interface-enhanced discrete element model (I-DEM) \cite{wu2024interface} further extends DEM by incorporating interfacial interlocking and associated failure mechanisms, facilitating accurate modeling of bio-inspired flexible protective structures. In spite of much light on the mechanics of beam-like substrates,  2D plate and shell-type systems are unique due to the nature of their cross-curvature couplings, emergent anisotropy and more complex scale sliding kinematics.

However, plate-like systems (Figure \ref{6-Fig1}(b)) remain much less studied with little progress in analytical
modeling. Intense scrutiny has ensued recently to address various aspects of these systems. These include novel computational frameworks to address existing gaps in simulation capabilities and shed new light on anisotropic behavior\cite{vernerey2014mechanics}. On similar lines, computational study of kinematics using conventional Lagrangian FE \cite{tatari2023bending} also highlights distinct anisotropic engagement modes based on the scale distribution and applied loads. These results strongly support experimental trends of these systems in existing literature \cite{zolotovsky2021fish}.  Novel fabrication and design techniques include the use of 3D printing for fabrication of optimized geometries \cite{martini2017comparative, connors2019bioinspired} and vibration-driven assembly methods for forming large-piece free-standing topologically interlocked panels using polyhedral building blocks \cite{bahmani2022vibration}. Furthermore, hybrid designs on armor-like systems that combine polymeric and viscous fluid-filled layer exhibit superior impact resistance \cite{du2023low}. These fabrication methods can result in novel and renewed applications such as instability suppression \cite{vernerey2014skin}, flexibility-strength combinations \cite{funk2015bioinspired}.

These studies of 2D biomimetic structures reveal the fundamentally different behavior of 2D structures from their 1D counterparts. Of special interest are biaxial asymmetries and cross-curvature couplings that get amplified when plate-like bending is taken into account. A thorough understanding of these effects requires quantification of salient features of the system and their relationship to each other. Yet, there is a gap in our understanding due to the lack of accurate and reliable analytical models to understand these 2D systems. In this paper, we develop, for the first time, an analytical model of the 2D scale-covered plate and its subsequent influence on the resultant nonlinear elasticity of the substrate in bending deformation. The scales are assumed to be rigid and the formulation is developed for both synclastic (Figure \ref{6-Fig1}(e)) and anti-clastic (Figure \ref{6-Fig1}(f)) deformation modes. Our model validates intuition and at the same time opens a new scientific understanding of the nature of anisotropy and non-linearity present in 2D scale-covered plates subjected to bi-directional bending. The kinematic formulation is further extended for the locking analysis of the plate along both longitudinal and lateral directions of the plate. To illustrate the anisotropic behavior of the scale-covered plate structure, motivational 3-point bending tests are also carried out.

The paper is organized into the following sections: in section \ref{PlateKinematics}, we present the kinematics model of the scale-covered plate for both synclastic and anti-clastic bending deformations. Section \ref{Mechanics} includes the detailed derivation on the mechanics of the plate problem leading to the analytical expressions of moment-curvature behavior. In section \ref{FEM} of the paper, we discuss the FE formulation of the problem to verify the analytical model. In section \ref{Sec: motivational experiment}, we describe the experimental procedure for conducting 3-point bending experiments on the scale-covered plate. In section \ref{Results}, we describe and discuss in detail, the obtained results showing the nonlinear and anisotropic behavior of these systems. Finally, we conclude the paper by highlighting the novelty, key findings, and potential future work in Section \ref{Conclusion}.

%%%%%%%%%%%%%%%%%%%%%%%%%%%%%%%%%%%%%%%%%%%%%%%%%%
\begin{figure} [t]
\begin{center}
\includegraphics[scale = 0.64]{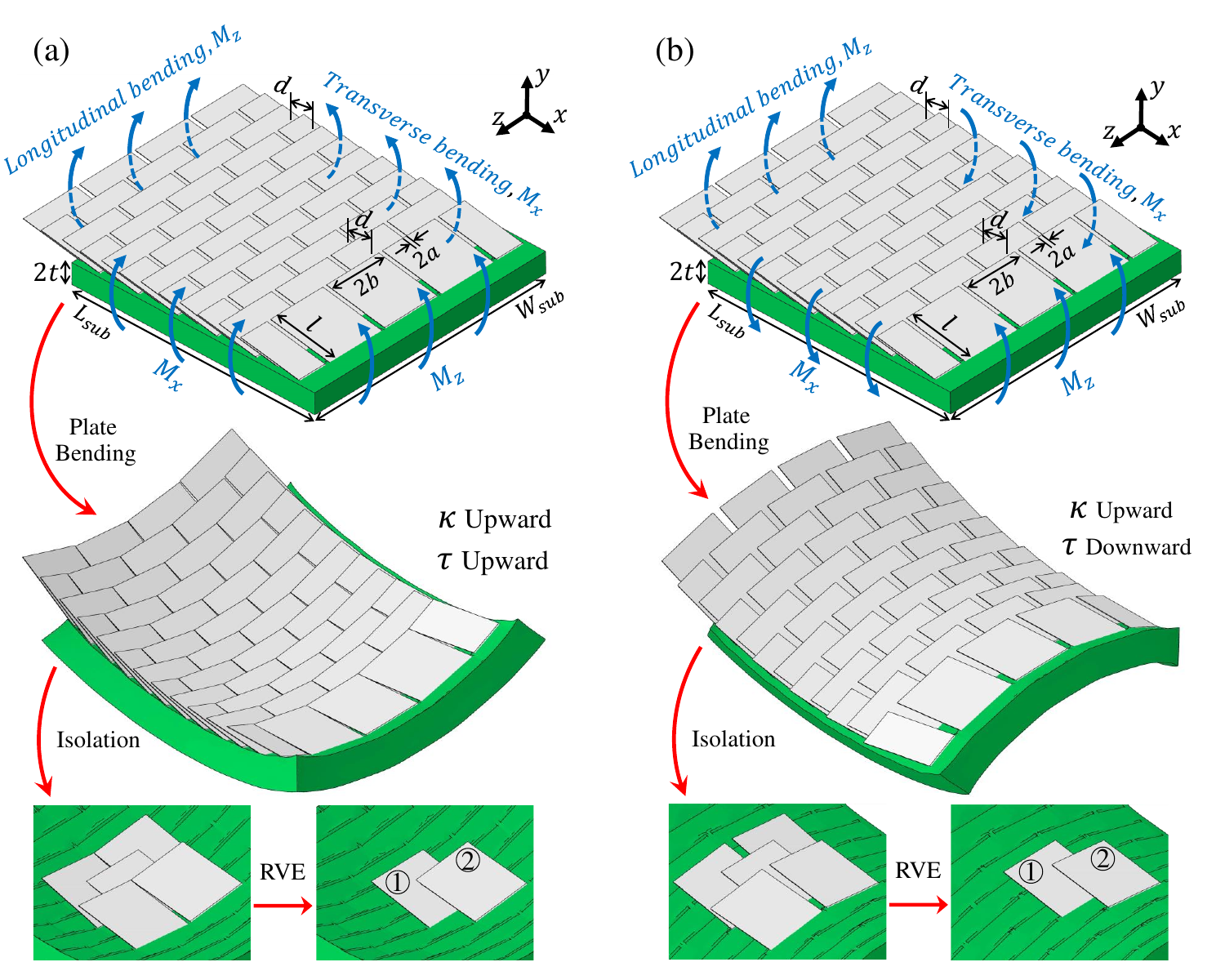}
\end{center}
\vspace{-15pt}
\caption{Scale-covered plate with the defined geometrical parameters and isolation of central scale and neighboring scales to demonstrate representative volume element (RVE) under two loading cases: (a) Synclastic deformation: both curvatures are upward ($\kappa>0$ and $\tau>0$). (b) Anticlastic deformation: longitudinal curvature is upward ($\kappa>0$), and transverse curvature is downward ($\tau<0$).}
\label{6-Fig2}
\end{figure}
%%%%%%%%%%%%%%%%%%%%%%%%%%%%%%%%%%%%%%%%%%%%%%%%%%

% \section{Experimental Testing}
% \label{Experiment}

\section{Contact Kinematics of Scale Sliding}
%\label{PlateBending}
\label{PlateKinematics}
In this section, the analytical expression is developed for the kinematics response of a deformable plate with scales covering its surface as shown in Figure \ref{6-Fig2}. The scales are partially embedded on the top surface of the plate with a staggered arrangement. Without loss of generality, we consider the deformable plate as a square shape with length and width of the substrate  $W_{sub}$ and $L_{sub}$, respectively (where, $W_{sub}=L_{sub}$), and thickness of the plate is $2t$. Bending in the longitudinal direction (along $x$-axis) is primarily assumed to happen due to the applied bending load $M_z$ in the planes perpendicular to $x$-axis. Similarly, bending in the transverse direction (along $z$-axis) of the plate is primarily assumed to occur due to the applied bending load $M_x$ in the planes perpendicular to $z$-axis, as shown in Figure \ref{6-Fig2}. The exposed length and the width of scales are $l$ and $2b$, respectively. The longitudinal and transverse distances between the scales are $d$ and $2a$, respectively. The dimensionless geometrical parameters $\eta=l/d$, $\beta=b/d$, $\lambda=t/d$, and $\delta=a/d$, can be defined as the overlap ratio, dimensionless scale width, dimensionless substrate thickness, and dimensionless clearance between the scales, respectively. The curvatures in the $x$-axis are called $\kappa$, which is the change rate of the slope angle $\psi$ with respect to the arc length in the longitudinal direction ($x$-axis). Also, the curvatures in the $z$-axis are $\tau$, which defines the rate of change of the slope angle $\omega$ with respect to the arc lengths in the transverse direction ($z$-axis). Two different load cases as shown in Figures \ref{6-Fig2} (a) and (b) are considered for analysis. In synclastic deformation, both curvatures are considered upward ($\kappa>0$ and $\tau>0$), and in anti-clastic deformation, the longitudinal curvature is kept upward ($\kappa>0$) and the transverse curvature is downward ($\tau<0$). The curvature equations can be approximated as $\kappa \approx\frac{\partial^2y}{\partial x^2}$ and $\tau \approx\frac{\partial^2y}{\partial z^2}$ (see \ref{AppendixA}).

\begin{figure} [t]
\begin{center}
\includegraphics[scale = 0.43]{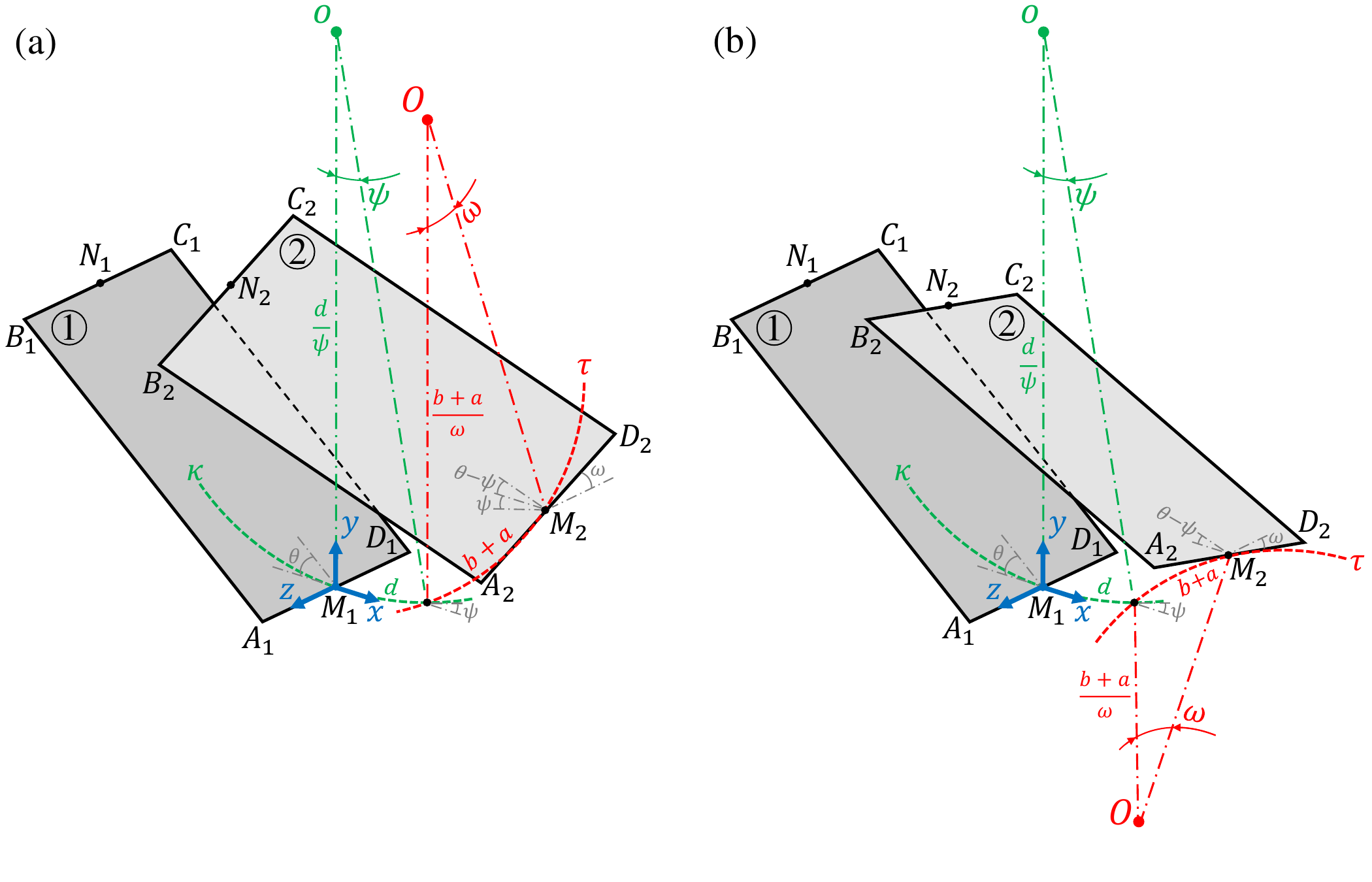}
\end{center}
\vspace{-15pt}
\caption{The schematic of the RVE geometrical configuration: (a) Both curvatures are upward ($\kappa>0$ and $\tau>0$), which means both local bending angles are positive ($\psi>0$ and $\omega>0$). (b) Longitudinal curvature is upward ($\kappa>0$) and the transverse curvature is downward ($\tau<0$), which means the local bending angles in $x$ direction is positive ($\psi>0$) and the local bending angles in $z$ direction is positive ($\omega<0$).}
%\vspace{-10pt}
\label{6-Fig3}
\end{figure}

Now, in the case of scale-covered plate, according to the pure bending in two plane directions, the periodicity assumption is a valid approximation for the scales engagement \cite{ghosh2014contact,ebrahimi2019tailorable,ebrahimi2020coulomb}. The periodicity assumption lets us isolate one scale (central scale) of the structure and its neighboring scales to model the behavior of the whole structure as per the principle of homogenization of periodic media, Figure  \ref{6-Fig2}. The central scale engages with 4 neighboring scales, and due to symmetry, the representative volume element (RVE) can be considered as the central scale and just one of the neighboring scales, as shown in Figures \ref{6-Fig2} (a) and (b) for two load cases including: (a) Both curvatures are upward ($\kappa>0$ and $\tau>0$), (b) The longitudinal curvature is upward ($\kappa>0$) and the transverse curvature is downward ($\tau<0$). The schematic of RVE is illustrated in Figure \ref{6-Fig3} (a) and (b) for both load cases. Thus although the underlying crystal is a typical Bravais-centered rectangle, the \textit{irreducible contact zone} (or the RVE) is just the central scale and one neighboring scale in this case. This RVE forms the basis of all of our subsequent structure-property calculations. 

In Figure \ref{6-Fig2} and Figure \ref{6-Fig3}, the central scales is denoted as ``1\textsuperscript{st} scale'' and shown as a rectangular plate $A_1B_1C_1D_1$. The neighboring scale is denoted as ``2\textsuperscript{nd} scale'' and shown as a rectangular plate $A_2B_2C_2D_2$. Due to the underlying deformation of the substrate with longitudinal curvature $\kappa$ and transverse curvature $\tau$, the 2\textsuperscript{nd} scale has been displaced and rotated as shown in Figure \ref{6-Fig3} for both synclastic and anti-clastic deformation. 

%%%%%%%%%%%%%%%%%%%%%%%%%%%%%%%%%%%%%%%%%%%%%%%%%%
%%%%%%%%%%%%%%%%%%%%%%%%%%%%%%%%%%%%%%%%%%%%%%%%%%

In synclastic deformation, the corner $B_2$ of 2\textsuperscript{nd} scale contacts with the top surface of 1\textsuperscript{st} scale. We place the coordinates $xyz$ on the midpoint of 1\textsuperscript{st} scale's width, $M_1$, as shown in Figure \ref{6-Fig3} (a). To find a contact criterion between 1\textsuperscript{st} scale and 2\textsuperscript{nd} scale, the 3D-equations of plane $A_1B_1C_1D_1$ and coordinate of point $B_2$ need to be established. To establish 3D-equations of plane $A_1B_1C_1D_1$, the coordinate of points $N_1$ and $D_1$ should be obtained, then the vectors ${\bm {M}}_{\bm 1}{\bm D}_{\bm 1}$ and ${\bm {M}}_{\bm 1}{\bm N}_{\bm 1}$ is obtained to calculate the normal vector of plane $A_1B_1C_1D_1$. Finally, the plate equation can be obtained by using the normal vector and coordinate of one of the points located in the plate. To find the coordinate of point $B_2$, located in the 2\textsuperscript{nd} scale, we start from the base point of 2\textsuperscript{nd} scale's plate, $M_2$. The $y$ coordinate of point $M_2$ is obtained from the deformed substrate's Equation (\ref{6-EqS8}) (see \ref{AppendixA}). The $x$ and $z$ coordinates of point $M_2$ are $d$ and $-(b+a)$, respectively. This means in the deformed state of the substrate, the local longitudinal and transverse arc lengths are equal to $d$ and $b+a$ respectively, as shown in Figure \ref{6-Fig3} (a), which leads to $\psi=\kappa d$ and $\omega=\tau(b+a)$ for the local bending angles. This leads to the local curvature radius in the longitudinal direction as $R_x=\frac{1}{\kappa}=\frac{d}{\psi}$, and the curvature radius in the transverse direction as $R_z=\frac{1}{\tau}=\frac{b+a}{\omega}$, as shown in Figure \ref{6-Fig3} (a).

Because the edge $A_2D_2$ is tangent to the substrate surface at the middle point $M_2$, the angle of this edge with the direction of $z$-axis is equal to $\omega$. According to this, the coordinate of points $A_2$ and $D_2$ can be found with respect to the point $M_2$. Also, the longitudinal symmetry line of 2\textsuperscript{nd} scale's plate, $M_2N_2$, has an angle equal to $\theta$ with respect to the tangent line of the substrate at point $M_2$ in the direction of $x$-axis. This means the line $M_2N_2$ has an angle equal to $\theta-\psi$ with respect to the direction of the $x$-axis. From this, the coordinate of point $N_2$ can be found with respect to the point $M_2$. Because line $A_2D_2$ is parallel to line $B_2C_2$, the locations of points $B_2$ and $C_2$, can be found with respect to the point $N_2$. After finding the location of point $B_2$, to satisfy the contact between this corner of 2\textsuperscript{nd} scale and the surface of 1\textsuperscript{st} scale, the location of point $B_2$ must satisfy the $A_1B_1C_1D_1$ plate equation. This leads to the following nonlinear kinematic relationship between scale inclination angle $\theta$, and the local bending angles $\psi$ and $\omega$:
\\
%%%%%%%%%%%%%%%%%%%%%%%%%%%%%%%%%%%%%%%%%%%%%%%%%%
\begin{equation}
\eta \sin \psi - \sin \theta - \cos \theta \Big(\frac{\psi}{2}+\big(\frac{\beta + \delta}{2}\big) \omega - \beta \sin \omega \Big) = 0. 
\label{6-Eq1}
\vspace{5pt}
\end{equation} 
%%%%%%%%%%%%%%%%%%%%%%%%%%%%%%%%%%%%%%%%%%%%%%%%%%

In synclastic deformation, instead of contacting the corner of 2\textsuperscript{nd} scale with the top surface of 1\textsuperscript{st} scale, the side edge $D_1C_1$ of 1\textsuperscript{st} scale contacts with the top edge $C_2B_2$ of 2\textsuperscript{nd} scale. In this case, as mentioned earlier the curvature in $z$ direction, $\tau$, and the local bending angle $\omega$ are negative ($\tau<0$ and $\omega<0$). The procedure for finding the coordinate of points $M_1$, $N_1$, and $D_1$ of the 1\textsuperscript{st} scale, and points $M_2$, $C_2$, and $B_2$ of the 2\textsuperscript{nd} scale are same as synclastic deformation, except that the curvature $\tau$ for Equation (\ref{6-EqS8}) should be considered as a negative value, because the transverse curvature is downward in this load case. After finding the coordinates of these points, the directional vector of lines $D_1C_1$ and $C_2B_2$ can be calculated, then the 3D-equations of these lines are obtained by using the directional vectors and one point on each line. To find the contact between these two lines, we solve their equations together as a system of equations, which yields the following nonlinear kinematic relationship between scale inclination angle $\theta$, and the local bending angles $\psi$ and $\omega$, which has a slight difference with respect to Equation (\ref{6-Eq1}):

%%%%%%%%%%%%%%%%%%%%%%%%%%%%%%%%%%%%%%%%%%%%%%%%%%
\begin{equation}
\eta \sin \psi - \sin \theta - \cos \theta \Big(\frac{\psi}{2}+\big(\frac{\beta + \delta}{2}\big) \omega - \delta \tan \omega \Big) = 0. 
\label{6-Eq2}
\end{equation} 
%%%%%%%%%%%%%%%%%%%%%%%%%%%%%%%%%%%%%%%%%%%%%%%%%%

The details of the mathematical derivation of Equation (\ref{6-Eq1}) and Equation (\ref{6-Eq2}) is given in sections \ref{AppendixB} and \ref{AppendixC}, respectively.
%%%%%%%%%%%%%%%%%%%%%%%%%%%%%%%%%%%%%%%%%%%%%%

\section{Mechanics of Plate Bending}
\label{Mechanics}
Similar to the pure bending and pure twisting cases \cite{ghosh2014contact,ebrahimi2019tailorable}, adding the rigid scales to the surface of the plate will provide additional stiffness response even before engagement. To account the effect of additional appreciable stiffness gain due to the scales inclusion into the substrate, two inclusion correction factors $C_{f,x}$ and $C_{f,z}$ are considered for two in-plane directions $x$- and $z$-axis, respectively. Also, after scales engagement each scale starts to rotate due to contact with other scales, and the rotation of the scale is resisted by the elastic substrate. Likewise 1-D bending and twisting cases \cite{ghosh2014contact,ebrahimi2019tailorable}, the substrate resistance is modeled as rotational springs, and the rotational spring constant $K_{\theta}^*$ is considered to account this effect \cite{ghosh2014contact,ebrahimi2019tailorable}. With all these considerations, we equate the work and energy at RVE level to obtain the moment-curvature relationship. 

The applied work due to the bending in longitudinal and transverse directions on the plate is $\int_{0}^{\kappa} M_z \,d\kappa$ and $\int_{0}^{\tau} M_x \,d\tau$, respectively. This work is absorbed by the plate deformation and scale rotation. The strain energy stored due to bending deformation of the plate is $\frac{D}{2} (C_{f,x} \kappa^2+ 2\nu \tau\kappa +C_{f,z} \tau^2) L_{sub} W_{sub}$ (see Equation (\ref{6-Eq6b}) of \ref{AppendixD}). Also, the energy absorbed by the scales after engagement is $\frac{1}{2} N_x N_z K_{\theta}^* (\theta - \theta _0)^2$; $N_x$ and $N_z$ are the number of scales along the $x$ and $z$- directions of the plate, respectively. Now, balancing all these work-energy expressions:

%%%%%%%%%%%%%%%%%%%%%%%%%%%%%%%%%%%%%%%%%
\begin{equation}
\begin{aligned}
&W_{\text{sub}}\int_{0}^{\kappa} M_z \,d\kappa + L_{\text{sub}}\int_{0}^{\tau} M_x \,d\tau \\
&= \frac{D}{2} (C_{f,x}\kappa^2+2\nu\kappa\tau+C_{f,z}\tau^2)L_{\text{sub}}W_{\text{sub}} + \frac{1}{2} N_x N_z K_{\theta}^* (\theta - \theta _0)^2 H(\theta - \theta _e).
\end{aligned}
\label{6-Eq3}
\end{equation}
%%%%%%%%%%%%%%%%%%%%%%%%%%%%%%%%%%%%%%%%%

%Here, $\theta_e$ is the engagement angle of scales. Substituting $W_{sub} = L_{sub}$ and $N_x = L_{sub}/d$ in Equation (\ref{6-Eq3}):   
%%%%%%%%%%%%%%%%%%%%%%%%%%%%%%%%%%%%%%%%%
%\begin{equation}
%\int_{0}^{\kappa} M_z \,d\kappa + \int_{0}^{\tau} M_x \,d\tau =\frac{D}{2} (C_{f,x}\kappa^2+2\nu\kappa\tau+C_{f,z}\tau^2)L_{sub} + \frac{1}{2} \frac{1}{d} N_z K_{\theta}^* (\theta - \theta _0)^2 H(\theta - \theta _e).
%\label{6-Eq4}
%\end{equation}
%%%%%%%%%%%%%%%%%%%%%%%%%%%%%%%%%%%%%%%%%

Here, $\theta_e$ is the engagement angle of scales. To obtain the moment-curvature relationship along longitudinal direction ($x$-axis) of the plate, we differentiate Equation (\ref{6-Eq3}) with respect to $\kappa$:

%%%%%%%%%%%%%%%%%%%%%%%%%%%%%%%%%%%%%%%%%%%%%%%%%%
%\vspace{-5pt}
%\begin{subequations}
\begin{equation}
W_{\text{sub}} M_z = D (C_{f,x} \kappa + \nu\tau) L_{\text{sub}}W_{\text{sub}} + K_{\theta}^*N_xN_z (\theta - \theta _0) \frac{\partial \theta}{\partial \kappa} H (\theta - \theta _e). \label{6-Eq4}
\end{equation} 
%%%%%%%%%%%%%%%%%%%%%%%%%%%%%%%%%%%%%%%%%%%%%%%%%%
 
%Where $D_B=\frac{E_B L_{sub}(2t)^3}{12(1-\nu^2)}$. 
Now, non-dimensionalizing Equation (\ref{6-Eq4}) by dividing $DW_{sub}$ and substituting the expression of $N_z = \frac{W_{sub}}{2(a+b)}$:

%%%%%%%%%%%%%%%%%%%%%%%%%%%%%%%%%%%%%%%%%%%%%%%%%%
%\vspace{-5pt}
%\begin{subequations}
\begin{equation}
\bar{M_z} = (C_{f,x} \kappa + \nu\tau) L_{\text{sub}} + \frac{K_{\theta}^*}{D} \frac{N_x}{2(a+b)} (\theta - \theta _0) \frac{\partial \theta}{\partial \kappa} H (\theta - \theta _e). \label{6-Eq5}
\end{equation} 
%%%%%%%%%%%%%%%%%%%%%%%%%%%%%%%%%%%%%%%%%%%%%%%%%%

Similarly, to obtain the moment-curvature relationship along the transverse direction ($z$-axis), we differentiate Equation (\ref{6-Eq3}) with respect to $\tau$, and, then non-dimensionalize  it by diving $DL_{sub}$ and substituting $N_x = \frac{L_{sub}}{d}$:

%%%%%%%%%%%%%%%%%%%%%%%%%%%%%%%%%%%%%%%%%
\begin{equation}
\bar{M_x} = (C_{f,z}\tau+\nu\kappa)W_{sub} + \frac{K_{\theta}^*}{D} \frac{N_z}{d} (\theta - \theta _0)\frac{\partial\theta}{\partial\tau} H(\theta - \theta _e).
\label{6-Eq6}
\end{equation}
%%%%%%%%%%%%%%%%%%%%%%%%%%%%%%%%%%%%%%%%%

Here, Equation (\ref{6-Eq5}) and Equation (\ref{6-Eq6}) are used to solve the moment-curvature response along $x$ and $z$-axis, respectively. From FE simulations, the constants $C_{f,x}$, and $C_{f,z}$ are calculated, and the detail of this procedure is given in \ref{AppendixD}. 

For rotational springs constant $K_{\theta}^*$, the scale width $2b$ is considered here with previously developed scaling law $K_{\theta}$ \cite{ghosh2014contact,ali2019bending}. Thus the final expression of $K_{\theta}^*$ will be:
%%%%%%%%%%%%%%%%%%%%%%%%%%%%%%%%%%%%%%%%%%%%%%%%%%%%%%

\begin{equation}
K_{\theta}^* =(2b)K_{\theta} = (2b)E_Bt_s^2 C_b (\frac{L}{t_s})^nf(\theta_0). \label{6-Eq7}
\end{equation}

%%%%%%%%%%%%%%%%%%%%%%%%%%%%%%%%%%%%%%%%%%%%%%%%%%%%%%
Here, $C_b$ = 0.66 and $n$ = 1.75 are dimensionless constant obtained from \cite{ghosh2014contact}. This equation of $K_{\theta}^*$ differs from the 1-D case of $K_{\theta}$ \cite{ghosh2014contact} is that when we consider scale as a rigid thin plate, scale width $2b$ also needs to be considered. And dimensionless angular function $f(\theta_0)$ $\approx$ 1 indicates negligible angular dependency in case of 2-D plate.

\section{Finite Element Analysis}
\label{FEM}

A Finite Element (FE) model is also developed to compare the numerical results with the results derived from the developed analytical model. The FE simulations are carried out in commercially available FE software ABAQUS/CAE (Dassault Syst`emes) keeping all the dimensions and loading conditions of the scale-covered plate as same as the analytical model.  The 3D deformable solids have been considered for the scales and the substrate. A length and width of $W_{sub}=L_{sub}$ = 64 mm are considered for the square plate substrate satisfying the periodicity and reducing the edge effect. The embedded length $L$ = 1 mm, half scale width $b$ = 7 mm, exposed length $l$ = 15 mm, scale thickness $t_s$ = 0.05 mm, and clearance between scale $a$ = 0.5 mm is considered which leads to $\eta = 3, \beta = 1.4$, $\delta = 0.1$, and $L/t_s$ = 20. An assembly of a staggered arrangement of 59 scales and the plate substrate has been considered, where the scales are partially embedded on the substrate’s top surface. The scales are oriented with the inclination angle of $\theta_0 = 5^o$ with respect to the substrate’s top surface in the longitudinal direction. Scales are assumed to be rigid and the substrate is considered as linear elastic material with the elastic modulus $E_B$ = 2.5 MPa and the Poisson’s ratio $\nu$ = 0.25.

To model the scales as rigid with respect to the deformable substrate, rigid body constraints are imposed on the scales’ geometry. A frictionless surface-to-surface contact interaction has been applied to the scales’ surfaces. The mesh convergence study is also carried out to discover a sufficient mesh size and density for different regions of the model, to obtain reliable numerical results. This leads to a total number of almost 354,000 elements in the mesh, including linear tetrahedral elements of type C3D4 and linear hexahedral elements of type C3D8. The top layer of the substrate is meshed with the tetrahedral elements, because the geometry is complex due to the scales inclusions into the substrate, but the other regions of the model are meshed with the hexahedral elements.

The bending mechanical loads are applied quasi-statically to the system through two static steps. To find the relationship between the scales inclination angle $\theta$ with the longitudinal bending angle $\psi$, while keeping the transverse bending angle $\omega$ is fixed at different values, or vice versa. For this purpose, in the first step in ABAQUS, the transverse rotational boundary conditions have been applied to both lateral sides of the substrate with reverse directions linearly increasing from zero to a desired value during the first step time. Then in the second step, the transverse rotational boundary conditions are kept fixed at the assigned value from the first step, and the longitudinal rotational boundary conditions have been applied to the both front and back sides of the substrate with reverse directions linearly increasing from zero. 

\section{Motivational Experiments}\label{Sec: motivational experiment}

%\subsection{Experimental analysis of the Plate}
In the experimental analysis, we have employed a multi-step manufacturing process consisting of creating a soft substrate and stiff scales commonly used in scales literature \cite{tatari2023bending}. To begin, we crafted a mold with dimensions of 220 $mm$ in length, 220 $mm$ in width, and 13 $mm$ in height. To create these mold and scales, we utilized a 3D printer (Ultimaker\textsuperscript{TM}) and employed Polylactic acid (PLA) as the printing material, which possesses an elastic modulus of approximately 3 $GPa$ ($E_{PLA}$). The soft substrate material was prepared by mixing silicon rubber (Dragon Skin\textsuperscript{TM} 10) with a curing agent in a 1:1 ratio. This mixture was then cast into the aforementioned mold. Once the substrate was adequately cured, it was carefully removed from the mold. Subsequently, the 3D-printed scales were affixed to it with adhesive, resulting in the final fabricated samples for our analysis. Thus, the dimensions of the square plate are 220 $mm$ in both length and width, with a thickness of 13 $mm$. The non-dimensional scale parameters of this sample are $L/t_s$ = 9, $\eta$ = 2.4, $\beta$ = 1.6, and $\theta_0 = 10^\circ$.

%%l=36, d=15, L=9, b=24, a =1.75, $\theta=15^0$, D=1mm

To illustrate the anisotropic behavior of the 2D scale-covered plate, we conducted a rigorous 3-point bending experiment. This experiment was carried out using the MTS Insight\textsuperscript{\textregistered} testing system, which enabled precise control of the displacement. The experiment was performed at a consistent cross-head speed of 1 $mm/s$, with the cross-head displacement ranging from 0 to 42 $mm$. We repeated each experiment three times to assess the consistency of the results. The objective of this 3-point bending experiment is to analyze the force-displacement response of the scale-covered plate in both longitudinal and transverse directions of the plate. When the plate is being bent in 3 point bending, a monoclastic deformation shape is made and to illustrate the anisotropy with the deformation shape we carried 3 points bending test along longitudinal and transverse direction of the plate as illustrated in the inset images of Figure \ref{6-Fig4}. This concave loading configuration has provided comprehensive data on the material's stiffness. Throughout the experiment, we continuously recorded the load cell values in correlation with the cross-head displacement. This allowed us to gain valuable insights into the anisotropic properties of the 2D plate, shedding light on how the scale-covered plate behaves in response to the applied bending in longitudinal and transverse directions separately.

\section{Results and Discussion}
\label{Results}
\subsection{Experimental Observations}
\vspace{10pt}

\begin{figure} [t]
\begin{center}
\includegraphics[scale = 0.6]{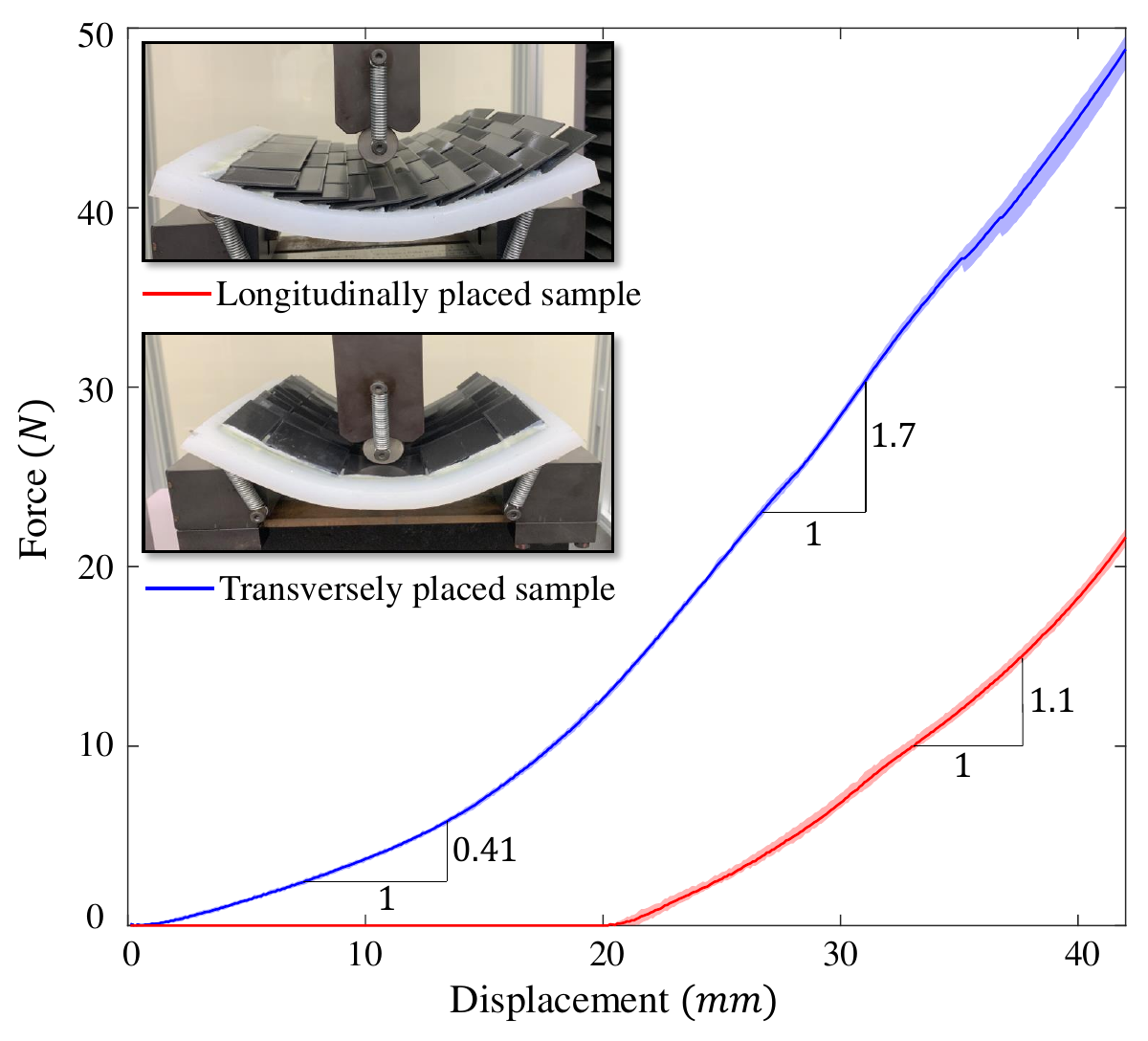}
\end{center}
\vspace{-15pt}
\caption{Force {\it vs} displacement plot for longitudinally and transversely placed 2D scale-covered plate in response to 3-point bending test. Here, $\eta$ = 2.4, $\beta$ = 1.6, and $\theta_0 = 10^\circ$.}
\label{6-Fig4}
\end{figure}

Figure \ref{6-Fig4} shows that the difference in bending rigidity between the longitudinal and transverse directions of the plate is significant. The figure considers the gravity induced sag in the samples. Interestingly, in the transversely placed configuration, the sag is minimal compared to the longitudinally placed configuration, suggesting greater stiffness even before the scales engage. This underscores the anisotropic behavior resulting from the inclusion of scales in the 2D plate, even prior to their engagement. This initial difference in instantaneous stiffness (tangent stiffness) is further amplified as loading commences, clearly indicating the intrinsic nonlinear anisotropy of this structure.

%%%%%%%%%%%%%%%%%%%%%%%%%%%%%%%%%%%%%%%%%%%%%%%%%%%%%%%%%%
\label{Fabrication}

%%%%%%%%%%%%%%%%%%%%%%%%%%%%%%%%%%%%%%%%%%%%%%%%%%%%%%%%%%%

\subsection{Contact Kinematics}
\vspace{10pt}

\begin{figure}[htbp]
\centering
\begin{tabular}{cc}
\includegraphics[scale = 0.75]{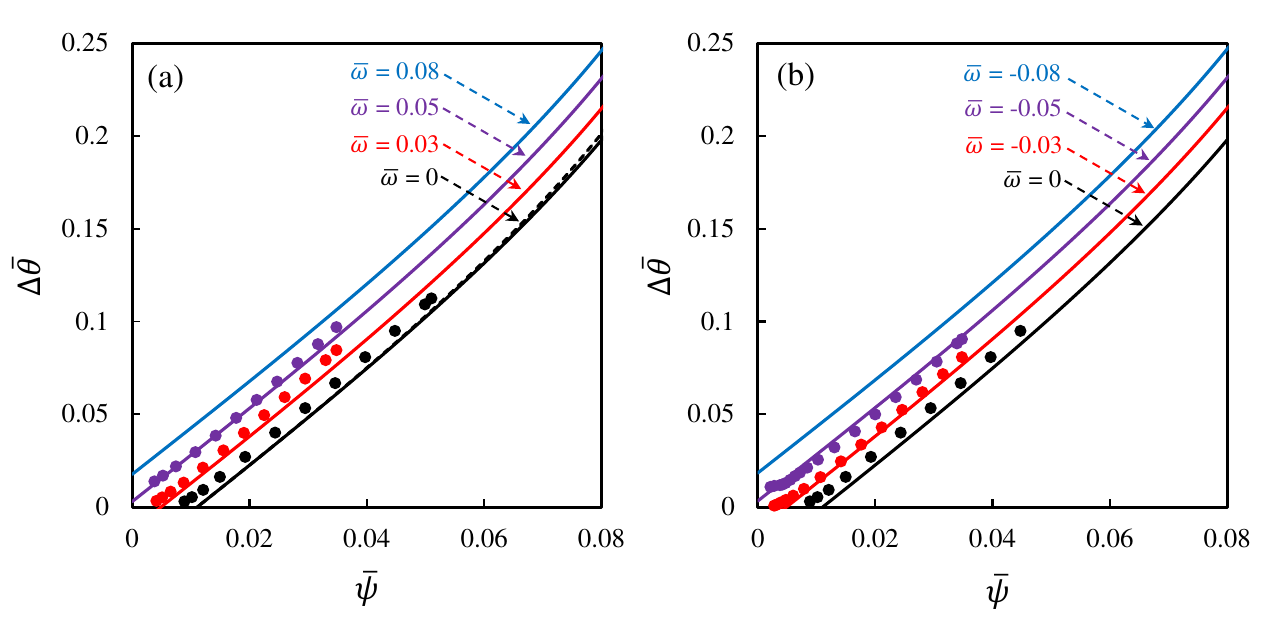} 
\end{tabular}
\caption{Plot of $\Delta\bar{\theta}$ = (($\theta-\theta_0$)/$\pi$) {\it vs} $\bar{\psi}$= $(\psi$/$\pi)$ of the scale-covered plate for different local transverse bending angle $\bar{\omega}$ for (a) synclastic, and (b) anti-clastic bending deformations with the given values of $\theta_0=5^\circ$, $\eta=3$, $\beta=1.4$, and $\delta=0.1$. Colored dot plots represent FE results for corresponding $\omega$. Black dashed line is obtained from Equation (1) of \cite{ghosh2016frictional}. }
 \label{6-Fig5}
\end{figure}

In Figures \ref{6-Fig5} (a)-(b), the normalized rotation $\Delta\bar{\theta}$ (=($\theta-\theta_0$)/$\pi$) of scale is plotted with respect to longitudinal curvature $\bar{\psi}$ (=$\psi$/$\pi$) for various values of transverse curvature $\bar{\omega}$ ($\omega$/$\pi$). In Figure \ref{6-Fig5} (a), results for synclastic ($\bar{\psi}$ $>$ 0 and $\bar{\omega}$ $>$ 0), and in Figure \ref{6-Fig5}(b), corresponding results related to anticlastic ($\bar{\psi}$ $>$ 0 and $\bar{\omega}$ $<$ 0) curvature is plotted. The solid dots indicate FE simulations for verification.  For both cases, it is clear that increasing the magnitude of transverse curvature leads to an earlier scales engagement, even for small deflections. Thus, there is a significant cross-curvature sensitivity in scale engagements. This behavior parallels earlier combined loading studies and confirms that cross-curvature sensitivity is a fundamental aspect of scaled system \cite{dharmavaram2022coupled}. More interestingly, in spite of the cross-curvature sensitivity of the engagement, transverse curvature did not seem to affect the $\Delta\bar{\theta}$-$\bar{\psi}$ slope, which would ultimately dictate the additional bending rigidity arising from scale contacts. Also the appearance of the $\Delta\bar{\theta}$-$\bar{\psi}$ plots remain the same for either synclastic or anticlastic deformation, which would indicate similar incremental rigidity from contacts in both deformation modes. This is an intriguing aspect of these systems where seeming asymmetry is nested within symmetries of other associated quantities. The colored dots representing FE results show excellent agreement with the analytical results. To compare the results with those available in the literature, we compared the results obtained from Equation (\ref{6-Eq1}) with the results of Equation (1) of \cite{ghosh2016frictional} for $\bar{\omega} = 0$. Equation (1) of \cite{ghosh2016frictional} was developed for 1D case of beam bending and as plotted in Figure \ref{6-Fig5}(a), in absence of transverse bending ($\bar{\omega}=0$), the newly developed Equation (\ref{6-Eq1}) shows a perfect match with Equation (1) of \cite{ghosh2016frictional} which also further reinforces the accuracy of the developed contact kinematics.

%%%%%%%%%%%%%%%%%%%%%%%%%%%%%%%%%%%%%%%%%%%%%%%%%%
%%%%%%%%%%%%%%%%%%%%%%%%%%%%%%%%%%%%%%%%%%%%%%%%%%
%%%%%%%%%%%%%%%%%%%%%%%%%%%%%%%%%%%%%%%%%%%%%%%%%%
\begin{figure} [htbp]
\centering
\begin{tabular}{cc}
\includegraphics[scale = 0.75]{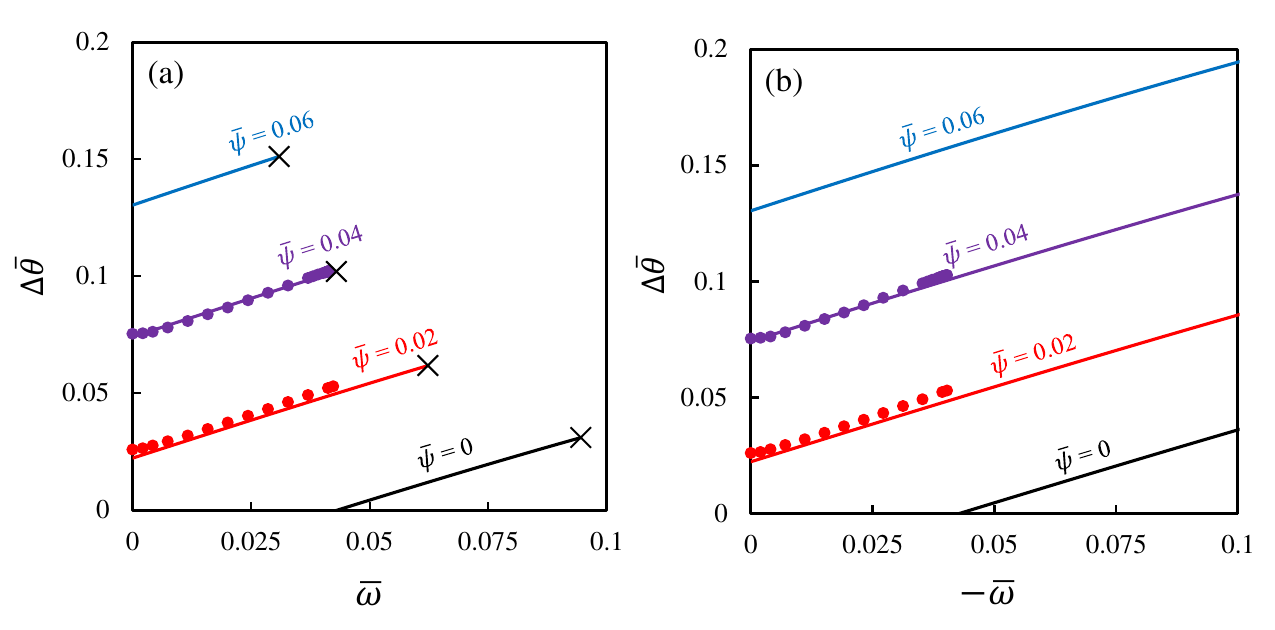} &
\end{tabular}
\caption{Plot of $\Delta\bar{\theta}$ = (($\theta-\theta_0$)/$\pi$) {\it vs} $\bar{\omega}$= $(\omega$/$\pi)$ of the scale-covered plate for different positive local longitudinal bending angle $\bar{\psi}$ for (a) synclastic, and (b) anti-clastic bending deformation with the given values of $\theta_0=5^\circ$, $\eta=3$, and $\beta= 1.4$. Colored dot lines represent FE results for corresponding $\psi$. ($\times$) indicates the lateral locking position of the plate.}
\label{6-Fig6}
\end{figure}
%%%%%%%%%%%%%%%%%%%%%%%%%%%%%%%%%%%%%%%%%%%%%%%%%%

%\vspace{10pt}

We now look at the cross-sensitivity of transverse curvature with respect to induced longitudinal curvature, $\bar{\psi}$. Figure \ref{6-Fig6} illustrates scale rotation $\Delta\bar{\theta}$ (local) variation with the global transverse bending curvature $\bar{\omega}$, across global longitudinal bending curvatures $\bar{\psi}$. To get these results, the biomimetic plate is first subjected to a longitudinal bending curvature $\bar{\psi}$, and then held constant, and then transverse curvatures $\bar{\omega} >0$ for synclastic and $\bar{\omega}<0$ for anticlastic curvature is applied. Four different values of $\bar{\psi}$ are considered from 0 to 0.06 with equal intervals. When the plate undergoes lateral bending, it reaches a specific curvature at which scales touch each other and become laterally locked, restricting further lateral deformation. As the deformation shapes shown in Figures \ref{6-Fig1} and \ref{6-Fig2}, indicate that lateral locking is achievable only in synclastic deformation; in the case of anti-clastic curvatures, lateral locking does not happen (The details of lateral locking calculations are given in \ref{AppendixB}). In both Figures \ref{6-Fig6}(a) and (b), an increase in the magnitude of longitudinal bending curvature increases the rotation of scale in either direction (synclastic and anti-clastic) without any significant effect on the slope of the $\Delta\bar{\theta}$ vs $\bar{\omega}$ dependence. This trend mirrors the previous results for $\Delta\bar{\theta}$ vs $\bar{\psi}$, Figures \ref{6-Fig6}(a) and (b). The $\Delta\bar{\theta}$-$\bar{\omega}$ curves show linear dependence for any cross longitudinal curvature value for both synclastic or anti-clastic curvature. Interestingly, no engagement of scales happens with transverse bending in the absence of longitudinal bending ($\bar\psi$ = 0). But when longitudinal bending is present even in a relatively small amount, scale engagement will occur with transverse bending. In synclastic deformation (Figure \ref{6-Fig6}(a)), locking curvature in the transverse direction of the plate is found to decrease with imposed longitudinal curvature $\bar{\psi}$. This means that imposed cross-curvature accelerates the transverse mode of locking, which also agrees with our intuition. The dotted colored lines represent FE results indicating an excellent agreement with the developed analytical model for both synclastic and anti-clastic deformations.

\subsection{Locking Behaviour of the Plate}
\vspace{10pt}

\begin{figure} [hbt!]%[htbp]
\begin{center}
\includegraphics[scale = 0.48]{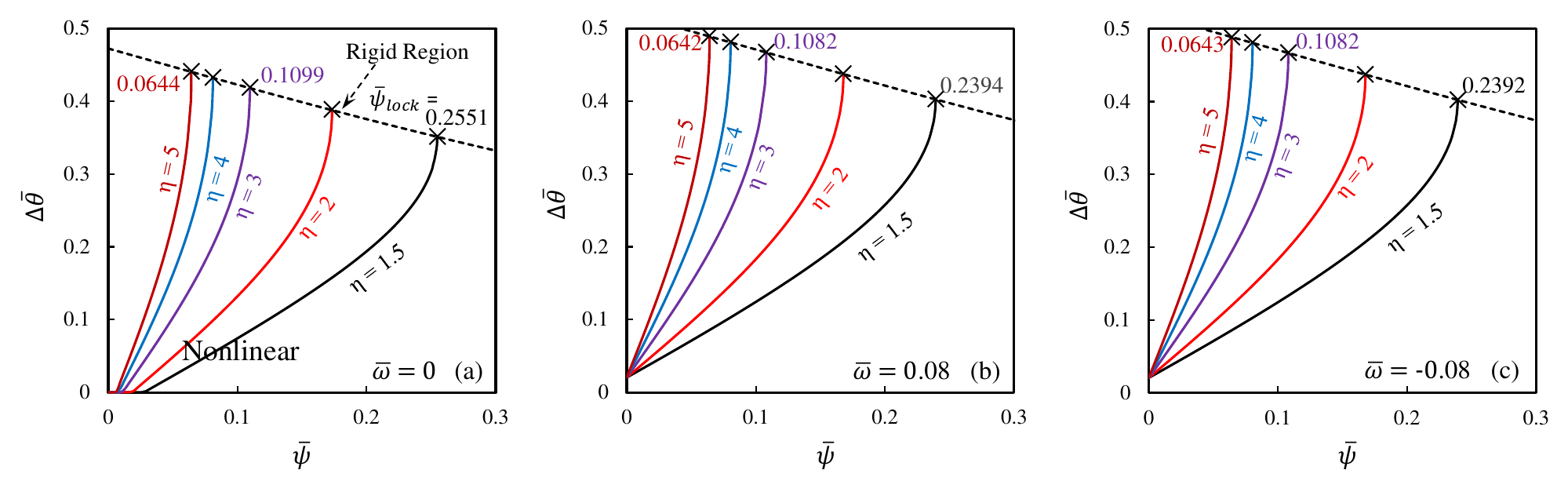}
\end{center}
\caption
{\label{6-Fig7} $\Delta\bar{\theta}$ {\it vs} $\bar{\psi}$ distribution with various $\eta$ when (a) $\bar{\omega}$ = 0, (b) $\bar{\omega}$ = 0.08, (c) $\bar{\omega}$ = -0.08 with $\bar{\psi}>0$. The value of $\bar{\psi}_{lock}$ $(\psi_{lock}/\pi)$ for corresponding $\eta$ is also shown. Here, $\theta_0=5^\circ$, and $\beta=1.4$.}
\end{figure}

\begin{figure}[htbp]
 \centering
 \begin{tabular}{cc}
 \includegraphics[scale = 0.7]{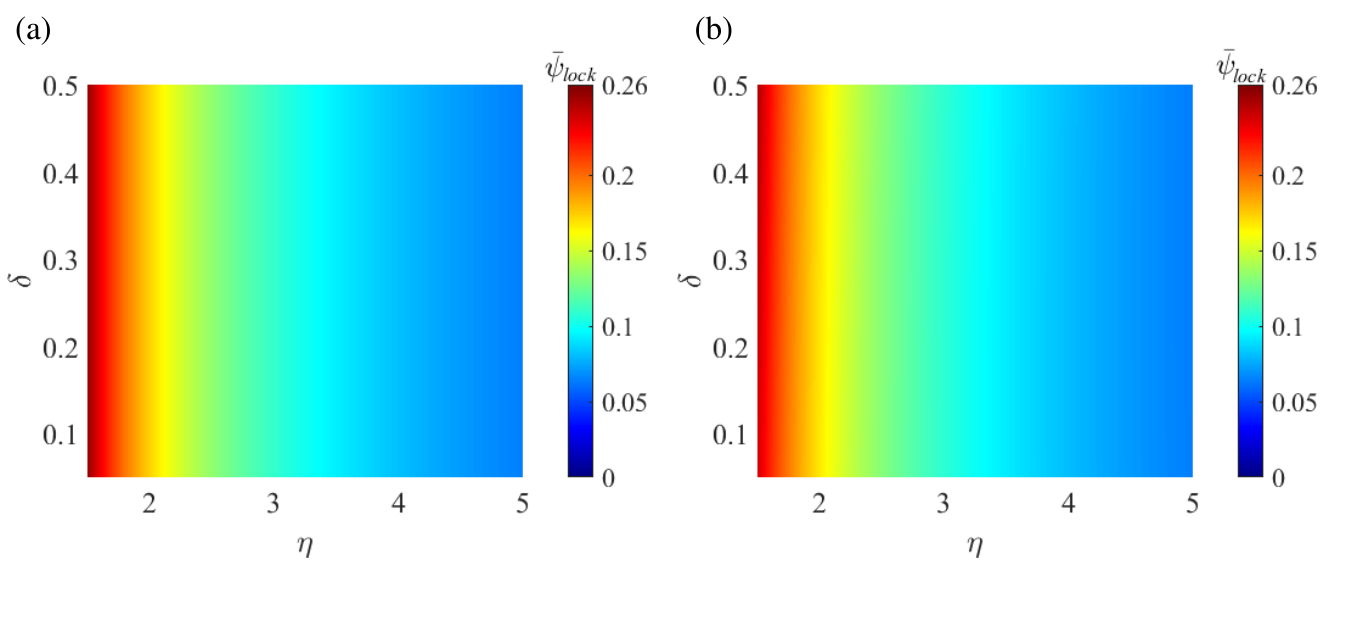} &
\end{tabular}
\caption{2D contour plots illustrate the locking response $\bar{\psi}_{lock}$ as a function of the increasing values of $\eta$ and $\delta$ resulting from loading at different local angular values: (a) $\bar{\omega} = 0$, and (b) $\bar{\omega}=0.08$ $(rad)$}
 \label{6-Fig8}
\end{figure}

\begin{figure} [hbt!]%[htbp]
\begin{center}
\includegraphics[scale = 0.48]{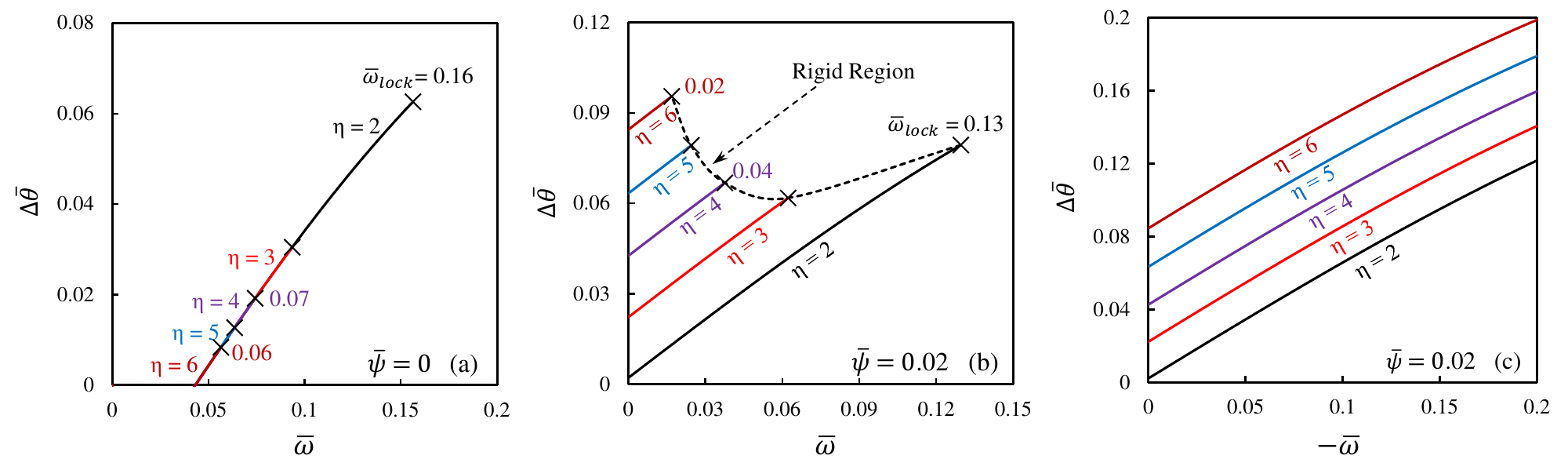}
\end{center}
\caption
{\label{6-Fig9} $\Delta\bar{\theta}$ {\it vs} $\bar{\omega}$ plot with various $\eta$ when (a) $\bar{\psi}$ = 0, (b) $\bar{\psi}$ = 0.02 and $\bar{\omega} > 0$, (c) $\bar{\psi} = 0.02$ and $\bar{\omega} < 0$. The value of $\bar{\omega}_{lock}$ $(\omega_{lock}/\pi)$  for corresponding $\eta$ is also shown. Here, $\theta_0=5^\circ$, and $\beta=1.4$.}
\end{figure}

\begin{figure} [hbt!]%[htbp]
\begin{center}
\includegraphics[scale = 0.43]{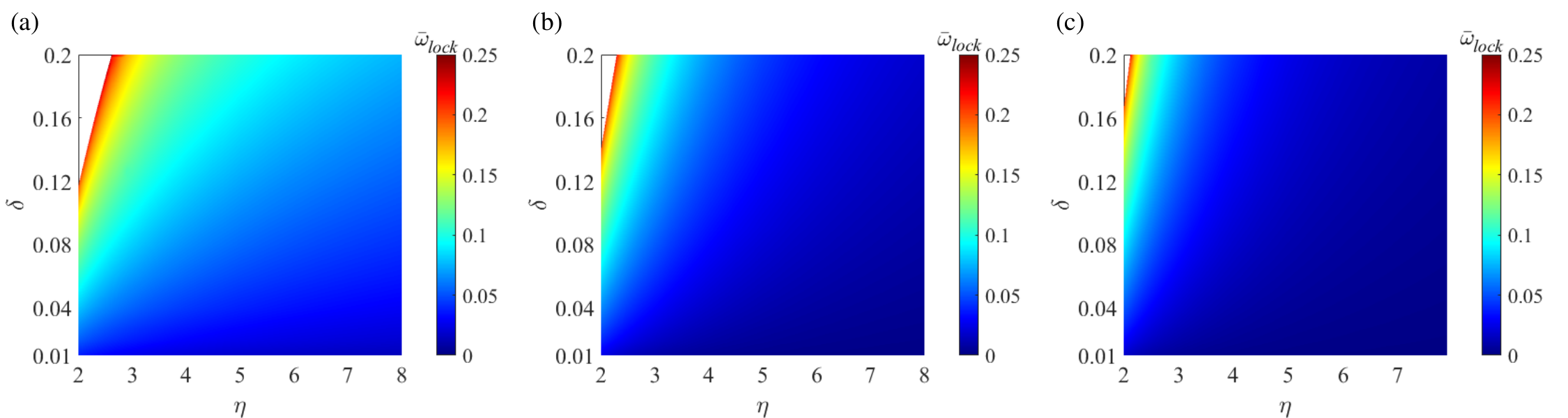}
\end{center}
\caption
{\label{6-Fig10} 2D contour plots illustrate the locking response ($\bar{\omega}_{lock}$) as a function of the increasing values of $\eta$ and $\delta$ resulting from loading at different local angular values: (a) $\bar{\psi} = 0$, (b) $\bar{\psi} =0.03$, and (c) $\bar{\psi} =0.05$.}
\end{figure}

We further explore the locking behavior alluded to in the previous section. Locking or jamming behavior of scaled substrates in bending, torsion, and combined loading has been a hallmark of biomimetic scale \cite{ebrahimi2019tailorable,ghosh2014contact}. At locking curvature, even without scale friction, the scales reach a kinematically forbidden state wherein they can no longer slide. In reality, the deformation in that case transitions from the substrate bending to the bending of the stiff scales leading to a sudden increase in stiffness. The kinematics of the scale sliding can provide critical clues on the extent, arrival, and nature of locking behavior.  In 1D substrates, locking is dictated by the scale overlap ratio $\eta$, with higher overlaps leading to earlier locking. In the 2D case, we once again notice similar behavior, Figure \ref{6-Fig7}, with applied longitudinal curvature when the transverse curvature is held at a constant value. Like the 1D case, for all values of overlap ratio, we observe three distinct bending regions – linear before engagement, nonlinear post-engagement, and finally a locked configuration. In spite of these broad similarities, plate geometry has a clear cross-curvature dependence of engagement and locking values, Figures \ref{6-Fig7}(a)-(c), with increased transverse curvature $\bar{\omega}$ in either direction (synclastic or anticlastic) advancing the locking curvature in the longitudinal direction. However, this advancement diminishes at higher overlap ratios, and surprisingly, no significant variation with positive (synclastic) and negative (anti-clastic) imposed transverse curvature on longitudinal locking is observed in Figures \ref{6-Fig7}(b) and (c). In addition to overlap ratios, the effect of transverse interscale clearance, defined by $\delta$ on the longitudinal locking is also of interest as it is the second lattice variable for this case. Here we define this as a ratio of transverse to longitudinal gap ($\delta$=$l/d$). This behavior is shown in Figure \ref{6-Fig8}, which indicates that for a given $\delta$, an increase in $\eta$ leads to a lowering of locking curvature for both neutral case (no transverse curvature) and positive curvature. On the contrary, for a given overlap ratio, even with a substantially high $\bar{\omega}$, $\delta$ seems to have no effect on the longitudinal locking curvature. This independence is a clear sign of a different locking mechanism at play. Indeed, in the transverse direction, the locking mechanism is that of non-sliding interlock (pillar jamming). Hence, as soon as contact is established, instant locking occurs in that direction.  Moreover, if we consider a thought experiment where $\delta$ goes to zero, we will get back a 1D beam (wide plate/fat beam). We further explored the locking in the other direction – i.e transverse curvature lock with imposed longitudinal curvature $\bar{\psi}$ in Figures \ref{6-Fig9}(a)-(c). In this case, with no imposed longitudinal curvature, Figure \ref{6-Fig9}(a), the locking merely advances with $\eta$ while keeping the slope of the $\Delta\bar{\theta}$ vs $\bar{\omega}$ for various values of $\eta$ constant. The collinearity of $\Delta\bar{\theta}$ vs $\bar{\omega}$ plots with various $\eta$ indicates that, in the absence of longitudinal bending, with only transverse bending, no engagement of scales will occur till the end, where transverse locking is achieved at the locking curvature of the plate (such behavior of scale rotation is also reported in Figure \ref{6-Fig6}(a)). These collinear locking curves split open when scales are engaged with an imposed positive longitudinal curvature (synclastic), Figure, \ref{6-Fig9}(b) indicating engagement and sliding. This leads to a familiar differential locking behavior, with imposed cross-curvature ($\bar\psi = 0.02$) once again providing earlier scales engagement and lower locking curvatures (advance of lock). As expected, a negative transverse curvature ‘opens up’ the scales leading to no locking behavior, Figure \ref{6-Fig9}(c). In order to find the effect of clearance ratio and overlap ratio on the transverse locking behavior, we plot an analogous phase diagram (to Figure \ref{6-Fig8}) in Figure \ref{6-Fig10}. Here the behavior is quite different from longitudinal locking.  In this case, there is indeed a strong dependence of side clearance $\delta$ on locking. Clearly, for lower values of $\eta$, if side clearance is sufficiently large, locking of scales in the transverse direction can be avoided. This lock-free region is sharply separated from the locking region, especially for the smaller magnitude of imposed positive longitudinal curvatures (synclastic). A trend is unmistakable from Figures \ref{6-Fig10}(a)-(c), wherein higher imposed longitudinal curvatures reduce the lock-free zone, or in other words facilitate locking. It can also be seen in the lock region that, for the same $\eta$ and $\delta$, an imposed $\bar{\psi}$ significantly reduces the curvature at which lateral locking takes place.

%%%%%%%%%%%%%%%%%%%%%%%%%%%%%%%%%%%%%%%%%%%%%%%%%%
%%%%%%%%%%%%%%%%%%%%%%%%%%%%%%%%%%%%%%%%%%%%%%%%%

%%%%%%%%%%%%%%%%%%%%%%%%%%%%%%%%%%%%%%%%%%%%%%%%%%
%%%%%%%%%%%%%%%%%%%%%%%%%%%%%%%%%%%%%%%%%%%%%%%%%%

%%%%%%%%%%%%%%%%%%%%%%%%%%%%%%%%%%%%%%%%%%%%%%%%%%
%%%%%%%%%%%%%%%%%%%%%%%%%%%%%%%%%%%%%%%%%%%%%%%%%

%%%%%%%%%%%%%%%%%%%%%%%%%%%%%%%%%%%%%%%%%%%%%%%%%%
%%%%%%%%%%%%%%%%%%%%%%%%%%%%%%%%%%%%%%%%%%%%%%%%%

\subsection{Strain energy of the Plate}
\vspace{10pt}

\begin{figure} [htbp]
\begin{center}
\includegraphics[scale = 0.44]{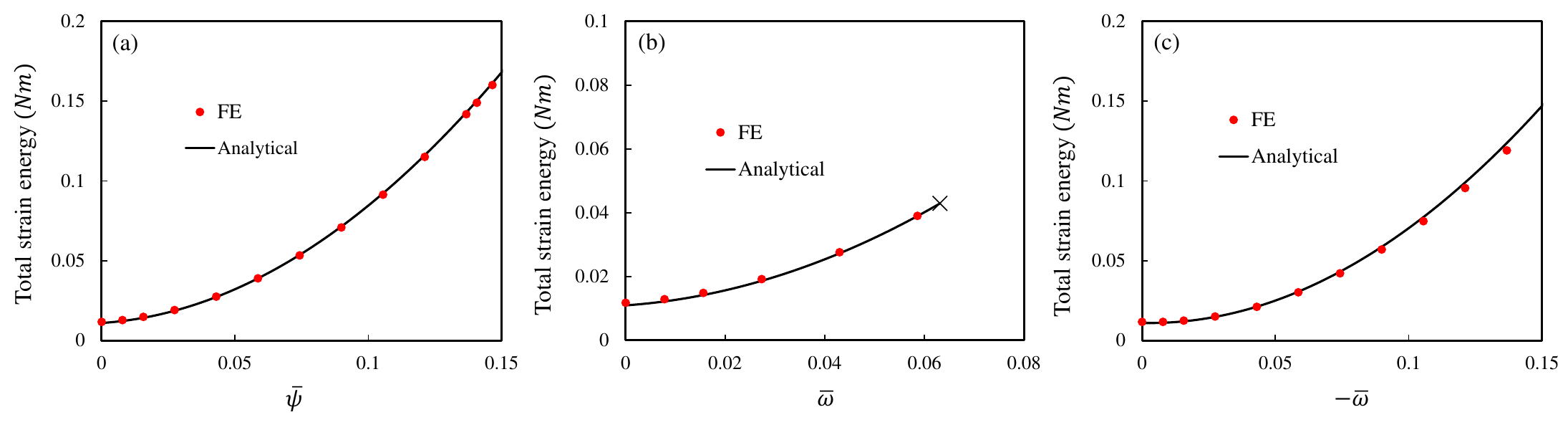}
\end{center}
\caption
{\label{6-Fig11} A comparison of analytical and FE results for total strain energy distribution with bending curvature of the plate when: (a) $\bar{\omega}$ = 0.03 and $\bar{\psi}>0$, (b) $\bar{\psi}$= 0.02 and $\bar{\omega}>0$, and (c) $\bar{\psi}$ = 0.02 and $\bar{\omega}<0$. Here, $\theta_0$ = 5$^\circ$, $\eta$ = 3, and $\beta$ = 1.4. ($\times$) in Figure 11(b) indicates the lateral locking position of the plate.}
\end{figure}

\begin{figure} [hbt!]%[htbp]
\begin{center}
\includegraphics[scale = 0.43]{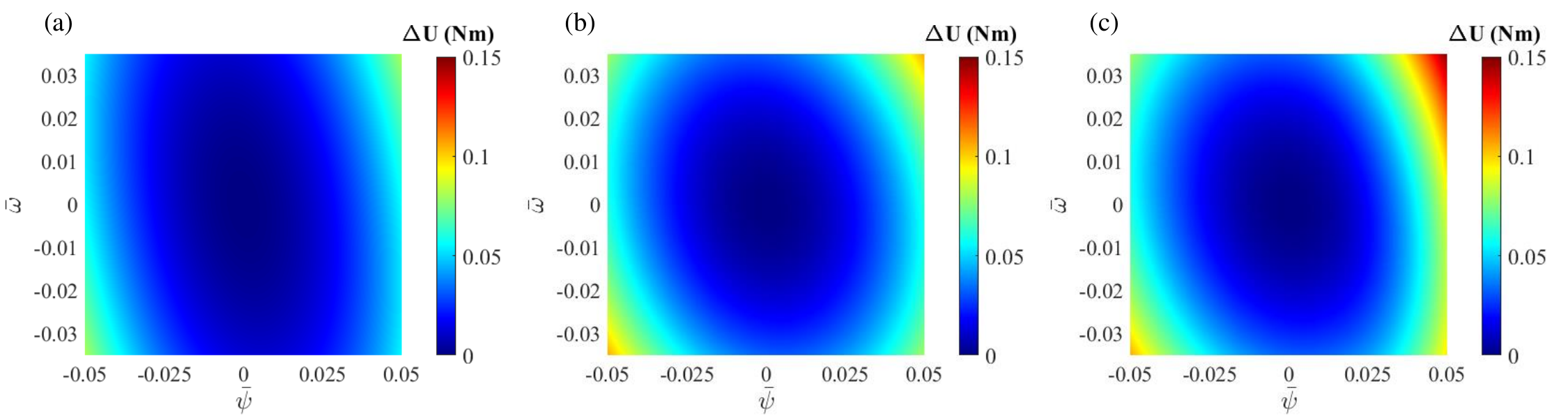}
\end{center}
\caption
{\label{6-Fig12} Strain energy distribution with longitudinal and transverse curvature for: (a) plain plate, (b) plate with rigid inclusion ($\eta = 0$), and (c) scale-covered plate with $\eta$ = 3. Here, $L_{sub} = W_{sub} = 64$ $(mm)$, $\beta$ = 1.4, and $\theta_0$ = 5$^\circ$.}
\end{figure}

Strain energy density of a bio-mimetic plate depends mainly on the nature of deformation, underlying plate material properties, and overlap ratio of scales. We first verified our expressions for the strain energy density of the biomimetic plate with FE simulations, Figure \ref{6-Fig11}. Next, we compare the analytical strain energy density map with imposed curvatures of a plain plate, Figure \ref{6-Fig12}(a), plate with only rigid inclusions, Figure \ref{6-Fig12}(b) (to mimic biomimetic plate before engagement), and finally plate with scales ($\eta$ = 3) sliding on each other, Figure \ref{6-Fig12}(c). Note that the scale inclusions break the diagonal symmetry in Figure \ref{6-Fig12}(a) due to their 'composite' effect on the sample. The presence of inclusions also rotates the energy landscape, Figure \ref{6-Fig12}(b). The anisotropy further accentuates with all symmetries virtually lost when scales sliding commences, Figure \ref{6-Fig12}(c), note the difference between upper-right and lower-left corners.

%%%%%%%%%%%%%%%%%%%%%%%%%%%%%%%%%%%%%%%%%%%%%%%%%%
%%%%%%%%%%%%%%%%%%%%%%%%%%%%%%%%%%%%%%%%%%%%%%%%%

%%%%%%%%%%%%%%%%%%%%%%%%%%%%%%%%%%%%%%%%%%%%%%%%%%
%%%%%%%%%%%%%%%%%%%%%%%%%%%%%%%%%%%%%%%%%%%%%%%%%

\subsection{Mechanics of the Plate}
\vspace{10pt}

\begin{figure}[htbp]
\begin{center}
\includegraphics[scale = 0.75]{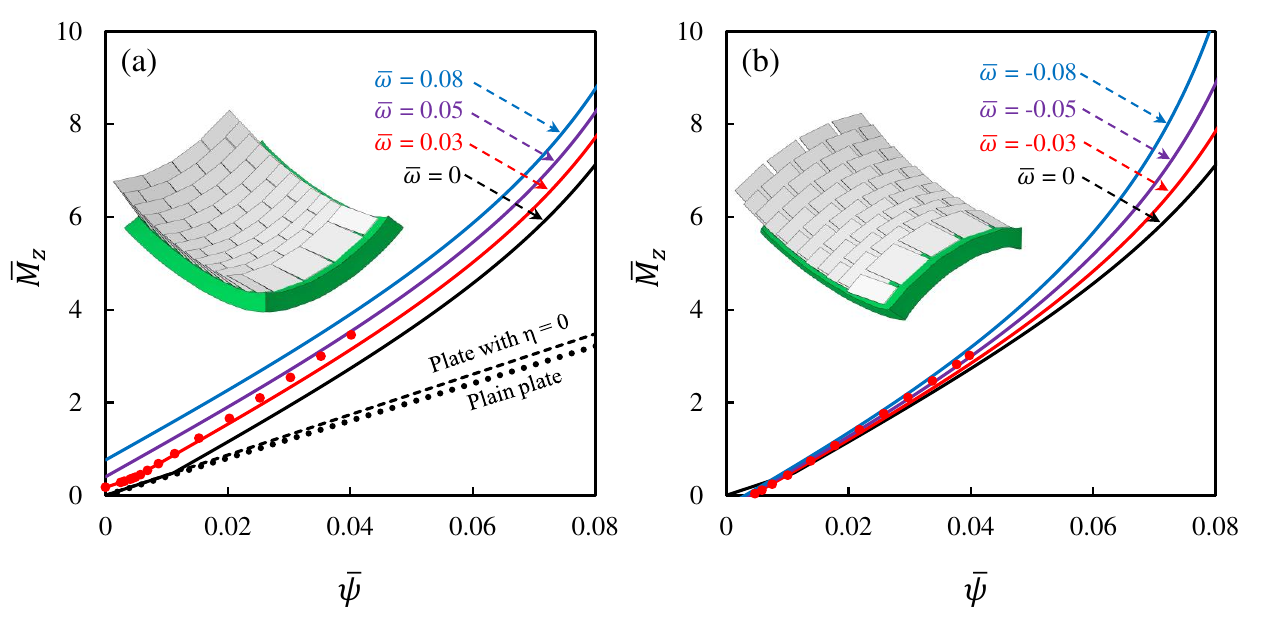}
\end{center}
\caption
{\label{6-Fig13} The normalized moment-curvature response of the scale-covered plate for different local transverse bending angle $\bar{\omega}$ for (a) synclastic, and (b) anti-clastic bending deformation with $\theta_0=5^\circ$, $\eta=3$, $\beta=1.4$, and $\delta=0.1$. Black dot and dashed lines (in Figure \ref{6-Fig13}(a)) represent plain plate and plates with $\eta$ = 0 results, respectively, and red dot line represents FE results for $\bar{\omega}$ = 0.03.}
\end{figure}

Figures \ref{6-Fig13}(a) and (b) respectively illustrate the non-dimensionalized moment–curvature ($\Bar{M_z}$) response for the longitudinal direction of the plate for synclastic and anti-clastic deformation ($\eta$ = 3) with four different initial values of transverse curvature $\bar{\omega}$. To analyze the effect of scale inclusion and scale engagement, a comparison with a plain plate and scale-embedded plate ($\eta$ = 0) is also presented. From the moment-curvature plot (Figure \ref{6-Fig13}(a)), the engagement of scales drastically increases the longitudinal moment. This moment increase happens earlier when a positive transverse curvature ($\bar{\omega}$) is imposed. This also indicates an earlier engagement of scales in the presence of transverse bending curvature keeping similarity with Figure \ref{6-Fig5}. The overall added rigidity is not very significant as slopes are relatively parallel for various positive transverse curvatures $\bar{\omega}$. The initial moment found in $\Bar{M_z}$ (when $\bar\psi$ = 0) is because of the Poission's effect of bending in the transverse direction. On the contrary, when an anticlastic transverse curvature is imposed, Figure \ref{6-Fig13}(b), the bending rigidities drastically increase with imposed transverse curvature which is completely in contrast of constant bending rigidies observed in the case of synclastic deformation. The red dotted lines shown in Figures \ref{6-Fig13} (a) and (b) are obtained from finite-element simulation for $\bar{\omega}$ = 0.03 and $\bar{\omega}$ = -0.03, respectively. As plotted in both figures, FE result shows excellent agreement with the analytical results of Equation (\ref{6-Eq5}) which verifies the proper modeling of the mechanics of the plate.

\begin{figure} [htbp]
\begin{center}
\includegraphics[scale = 0.74]{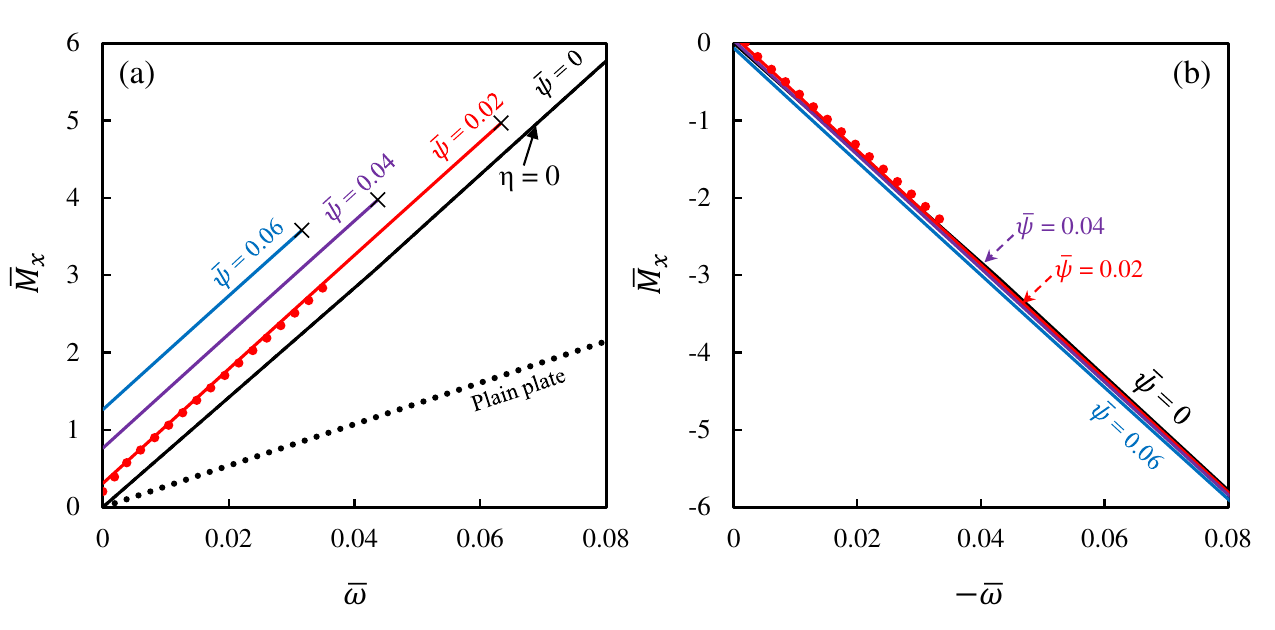}
\end{center}
\caption
{\label{6-Fig14} The normalized moment-curvature response for of the scale-covered plate for different positive local longitudinal bending angle $\bar{\psi}$ for (a) synclastic, and (b) anti-clastic bending deformation with $\theta_0=5^\circ$, $\eta=3$, $\beta=1.4$, and $\delta=0.1$. Black dotted and dashed lines (in Figure \ref{6-Fig14}(a)) represents results for plain plate and plates with $\eta$ = 0, respectively, and ($\times$) indicates the lateral locking position of the plate. Red dot lines represent FE results for $\bar{\psi}$ = 0.02.}
\end{figure}

%\vspace{10pt}
In Figures \ref{6-Fig14} (a) and (b), the normalized moment-curvature response along the transverse direction of the plate ($\Bar{M_x}$) is shown for both synclastic and anti-clastic deformation, respectively. Results are plotted for $\eta$ = 3 initially keeping the plate fixed with four different $\bar{\psi}$. A comparison of results are also shown with plain plate, and plate with rigid inclusions ($\eta$ = 0). In the case of synclastic deformation (Figure \ref{6-Fig14} (a)), the moment-curvature curve is plotted up to the lateral locking curvature of the plate. Unlike the longitudinal moment case, the transverse moment-curvature plots are linear till locking for all cross longitudinal curvature ($\bar{\psi}>0$) values. The slopes are nearly identical as well indicating relative independence of bending rigidity on cross-curvature effects. The plain plate and $\eta$ = 0 plots show that the stiffness gain due to the inclusion of scales is much more prominent in the lateral direction compared to the longitudinal direction. Also, a direct consequence of the kinematics of engagement, \textit{i.e.}, no engagement of scales with only transverse bending (see Figure~\ref{6-Fig9}(a)), is reflected in the associated collinearity of the moment-curvature plots for $\eta$ = 0 and 3, when $\bar\psi=0$. In contrast to synclastic bending, Figure 14(b), the anti-clastic deformation of the plate exhibits a negligible effect of $\bar{\psi}$ on the lateral moment response. The FE results plotted for $\eta$ = 3 also show good agreement with analytical results for both loading conditions. In conclusion, our discussions in subsections 6.4 and 6.5 show the deformation parameters governing the nonlinear anisotropic response of plate bending, especially the influence of cross curvatures on moment-curvature and their effects on the overall energy landscape.

%%%%%%%%%%%%%%%%%%%%%%%%%%%%%%%%%%%%%%%%%%%%%%%%%%
%%%%%%%%%%%%%%%%%%%%%%%%%%%%%%%%%%%%%%%%%%%%%%%%%
%%%%%%%%%%%%%%%%%%%%%%%%%%%%%%%%%%%%%%%%%%%%%%%%%%
%%%%%%%%%%%%%%%%%%%%%%%%%%%%%%%%%%%%%%%%%%%%%%%%%

\subsection{Effect of $\eta$ and $\delta$ on the Mechanics of the Plate}
\vspace{10pt}

\begin{figure} [htbp]
\begin{center}
\includegraphics[scale = 0.75]{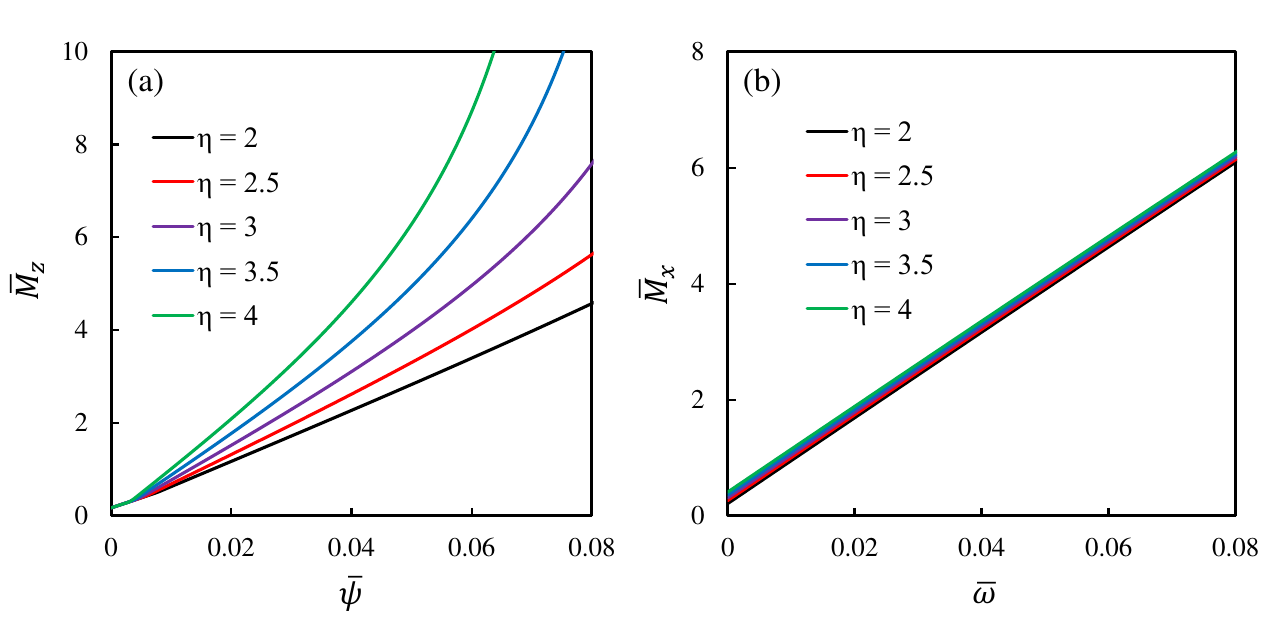}
\end{center}
\caption
{\label{6-Fig15} The normalized moment-curvature response for various $\eta$ for synclastic curvature along: (a) $x$-axis with $\bar{\omega}$ = 0.03 and $\bar{\psi}>0$, (b) $z$-axis with $\bar{\psi}$ = 0.02 and $\bar{\omega}>0$. Here, $\theta_0=5^\circ$, $\beta=1.4$, and $\delta=0.1$.}
\end{figure}

This section explores the effect of geometric quantities $\eta$ and $\delta$ in affecting the potential anisotropy of plates. Figures \ref{6-Fig15} (a) and (b) illustrate the effect of overlap ratio $\eta$ on the moment-curvature responses of the scale-covered plate in the corresponding longitudinal and transverse directions. Five different values of $\eta$ are considered from $\eta$ = 2 to 4, wherein the exposed length of the scale $l$ is varied while keeping $d$ constant. These results are plotted only for synclastic bending deformation where the applied moment in both in-plane directions is positive. In Figure \ref{6-Fig15}(a), initially the plate is kept fixed at $\bar{\omega} = 0.03$, then the longitudinal moment is plotted for increasing $\bar{\psi}$. Similarly, in Figure \ref{6-Fig15}(b), the transverse moment is plotted for increasing positive $\bar{\omega}$ while keeping the longitudinal curvature $\bar{\psi} = 0.02$ fixed. As we see in Figure \ref{6-Fig15}(a), the longitudinal rigidity gain after scales engagement of the plate is highly sensitive on $\eta$. For transverse-directional moment plotted in Figure \ref{6-Fig15}(b), the increasing $\eta$ has a negligible effect on the moment-curvature variation of the plate, which is completely in contrast with the longitudinal case, once again highlights the inherent anisotropy of this system. It is worth mentioning that, in anti-clastic loading conditions, the moment-curvature plots for both the longitudinal and transverse directions exhibit similar trends as observed in the synclastic case, but the figures are left out for the sake of brevity. Still, it is important to note the consistency of the moment-curvature responses between the two loading conditions.

%%%%%%%%%%%%%%%%%%%%%%%%%%%%%%%%%%%%%%%%%%%%%%%%%%
%%%%%%%%%%%%%%%%%%%%%%%%%%%%%%%%%%%%%%%%%%%%%%%%%%

%%%%%%%%%%%%%%%%%%%%%%%%%%%%%%%%%%%%%%%%%%%%%%%%%%
\begin{figure} [htbp]
\begin{center}
\includegraphics[scale = 0.75]{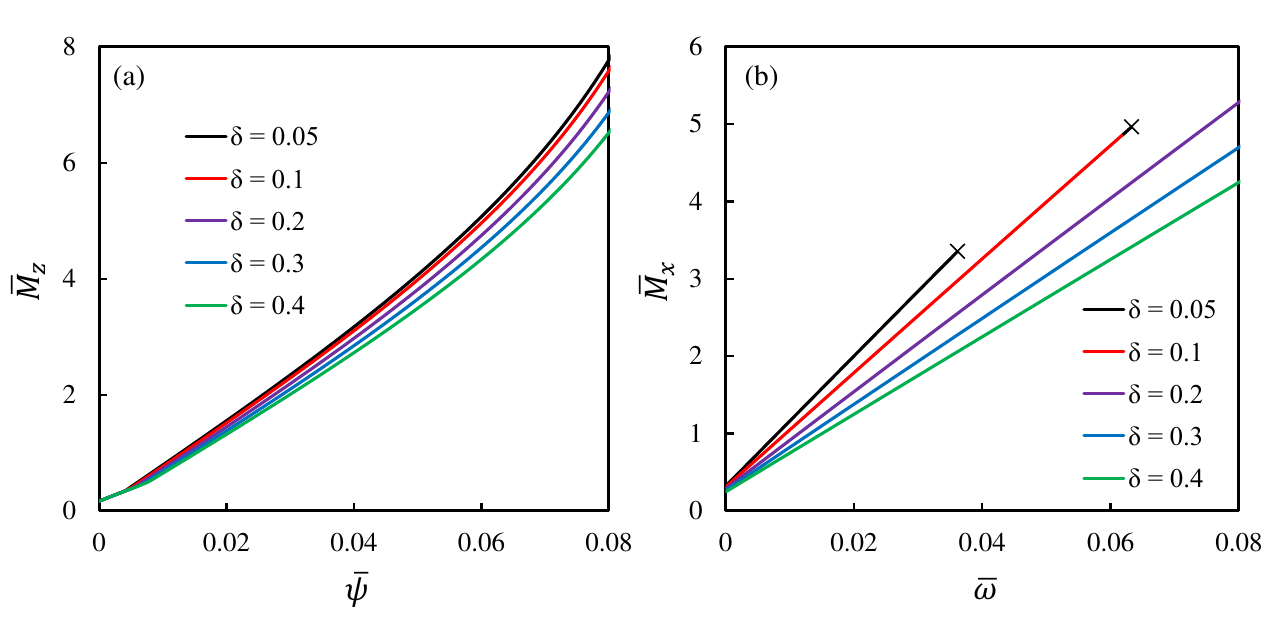}
\end{center}
\caption
{\label{6-Fig16} The normalized moment-curvature response for various $\delta$ for synclastic curvature along: (a) $x$-axis with $\bar{\omega}$ = 0.03 and $\bar{\psi}>0$, (b) $z$-axis with $\bar{\psi}$ = 0.02 and $\bar{\omega}>0$. Here, $\theta_0=5^\circ$, $\eta=3$, and $\beta=1.4$. ($\times$) in Figure  \ref{6-Fig16}(b) indicates the lateral locking position of the plate.}
\end{figure}
%%%%%%%%%%%%%%%%%%%%%%%%%%%%%%%%%%%%%%%%%%%%%%%%%%

%\vspace{10pt}
Figures \ref{6-Fig16} (a) and (b) demonstrate the effect of dimensionless clearance ratio $\delta$ on the moment-curvature response of the plate in both the longitudinal and transverse directions for synclastic deformation shape. To illustrate the effect of $\delta$ variations, half of the transverse distance between the scales $a$ is varied while keeping $d$ constant. Likewise, in the previous $\eta$ study, in this $\delta$ analysis as well, the plate is initially kept fixed at transverse curvature $\bar{\omega} = 0.03$; then, the longitudinal moment is plotted for increasing $\bar{\psi}$ (Figure \ref{6-Fig16}(a)). Similarly, in Figure \ref{6-Fig16}(b), transverse moment-curvature plot is illustrated for increasing positive $\bar{\omega}$ while longitudinal direction initially held fixed at $\bar{\psi} = 0.02$. An analysis of gradually increasing $\delta$ of the plate from 0.05 to 0.4 shows that the moment-curvature relationship of the plate along both directions are strongly influenced with the change in $\delta$, with bending rigidity decreasing as $\delta$ increases. It is also noticeable from Figures \ref{6-Fig16}(a) and (b) that $\delta$ variation is more significant in the transverse directional moment compared to the longitudinal direction of the plate. Thus, observing both figures, it can be said that with the increase of $\delta$, which corresponds to a decrease in $\beta$, the plate structures experience a significant reduction in longitudinal rigidity, and the rigidity reduction is even more pronounced in the transverse direction. In anti-clastic deformation, the effect of $\delta$ on the moment-curvature of the plate is almost as same as the one observed in the synclastic deformation case and is omitted for brevity.

\section{Conclusions}
\label{Conclusion}

We develop, for the first time, an analytical model for the nonlinear contact kinematics of biomimetic scales for plate-like substrates and quantifies the origins of nonlinear and anisotropic behavior of scale-covered plates. This study completes the pathway to extend the theoretical framework of biomimetic scale-covered systems from 1D to 2D structures. Here, the scales are considered as rigid rectangular plates, partially embedded with an inclination angle from the top surface of the elastic substrate. The scales are arranged with staggered arrangement (centered rectangular lattice) in two in-plane directions. Both synclastic and anticlastic curvatures of the plate are considered. We verified the model by comparing with finite-element simulations performed using commercially available software and beam-like configurations in literature. Our analysis shows highly intricate effects of scale engagement on the kinematics and mechanics of the system particularly on the nature of nonlinearity of bending, locked states, cross-curvature couplings, and emergent deformation-dependent anisotropy.  Although comprehensive, this model considers rectangular scales placed in a centered rectangular lattice. Other lattices are possible, but the overall technique revealed in this paper would stay the same. We also consider the underlying substrate linear elastic and behave according to Kirchhoff plate kinematics and the scales to be rigid. These are reasonable assumptions for thin, soft substrates with much stiffer scales where strains are low. FE validations are up to relatively small sliding angles due to inherent problems of the commercial FE codes in simulating large numbers of rigid contacts.  Despite these limitations, this work reveals the mechanistic origins of nonlinear and anisotropic behavior of biomimetic scale and other exoskeleton systems setting the ground for analyzing the effects of scale friction, fluid-scale interactions, and nonlinear dynamics. Such architecture-property relationships hold great potential for concerted computer-driven design and discovery of many real-world structural systems such as soft robotics, architectured structural metamaterials, protective armors and wearable materials, and multifunctional aerospace structures.

%%%%%%%%%%%%%%%%%%%%%%%%%%%%%%%%%%%%%%%
\section*{Declaration of competing interest}
The authors declare that they have no known competing financial interests or personal relationships that could have appeared
to influence the work reported in this paper.

%%%%%%%%%%%%%%%%%%%%%%%%%%%%%%%%%%%%%%%%%%%%%%%%%%%%%%%%%%%%%
\section*{Acknowledgement}
This work was supported by the United States National Science Foundation's Civil, Mechanical, and Manufacturing Innovation, CAREER Award \#1943886.

\appendix
\section{Plain Plate Deformation under 2D Bending} 
\label{AppendixA}

To determine the deformation shape of the underlying substrate, we initially considered the substrate (without any scales) as 2D thin plain plate subjected to bi-directional bending with synclastic and anti-clastic deformation modes. The curvature expressions along the longitudinal and transverse directions of the plain plate can be written as:

%%%%%%%%%%%%%%%%%%%%%%%%%%%%%%%%%%%%%%%%%%%%%%%%%%
%\vspace{-5pt}
\begin{subequations}
\begin{equation}
\kappa = \frac{\partial^2y/\partial x^2}{[1+(\partial y/\partial x)^2]^{3/2}}\approx\frac{\partial^2y}{\partial x^2}, \label{6-EqS1}
\end{equation} \vspace{-5pt}
\begin{equation}
\tau = \frac{\partial^2y/\partial z^2}{[1+(\partial y/\partial z)^2]^{3/2}}\approx\frac{\partial^2y}{\partial z^2}. \label{6-EqS2}
\end{equation} %\vspace{-16pt}
%\label{6-EqS1}
\end{subequations}
%\vspace{-5pt}
%%%%%%%%%%%%%%%%%%%%%%%%%%%%%%%%%%%%%%%%%%%%%%%%%%

According to classical plate theory or the Kirchhoff plate theory \cite{bauchau2009structural, reddy2006theory}, the moment-curvature response of the plate for two in-plane directions (similar to loading shown in Figure \ref{6-Fig2}) are as follows:

%%%%%%%%%%%%%%%%%%%%%%%%%%%%%%%%%%%%%%%%%%%%%%%%%%
%\vspace{-5pt}
\begin{subequations}
\begin{equation}
M_z = D (\frac{\partial^2y}{\partial x^2} + \nu\frac{\partial^2y}{\partial z^2}), \label{6-EqS3}
\end{equation} \vspace{-5pt}
\begin{equation}
M_x = D (\frac{\partial^2y}{\partial z^2} + \nu\frac{\partial^2y}{\partial x^2}). \label{6-EqS4}
\end{equation} \vspace{5pt}
%\label{6-EqS2}
\end{subequations}
%%%%%%%%%%%%%%%%%%%%%%%%%%%%%%%%%%%%%%%%%%%%%%%%%%

where $D=\frac{Eh^3}{12(1-\nu^2)}$, and $E$, $\nu$, and $h$ are the modulus, Poisson's ratio, and plate thickness, respectively. From Equations (\ref{6-EqS3}) and (\ref{6-EqS4}), the curvatures can be obtained as follows:
%%%%%%%%%%%%%%%%%%%%%%%%%%%%%%%%%%%%%%%%%%%%%%%%%%
%\vspace{-5pt}
\begin{subequations}
\begin{equation}
\kappa = \frac{\partial^2y}{\partial x^2} = \frac{M_z-\nu M_x}{D(1-\nu^2)}, \label{6-EqS5}
\end{equation} \vspace{-5pt}
\begin{equation}
\tau = \frac{\partial^2y}{\partial z^2} = \frac{M_x-\nu M_z}{D(1-\nu^2)}. \label{6-EqS6}
\end{equation} %\vspace{-16pt}
%\label{6-EqS3}
\end{subequations}
%%%%%%%%%%%%%%%%%%%%%%%%%%%%%%%%%%%%%%%%%%%%%%%%%%

According to equilibrium, the moments $M_x$ and $M_z$ are applied to all points constantly throughout the plate. Therefore, based on the moment-curvature Equations (\ref{6-EqS3}) and (\ref{6-EqS4}), the deformed plate equation under bending loads $M_x$ and $M_z$ is as follows:
%%%%%%%%%%%%%%%%%%%%%%%%%%%%%%%%%%%%%%%%%%%%%%%%%%
\vspace{8pt}
\begin{equation}
y = \frac{1}{2}\frac{M_z-\nu M_x}{D(1-\nu^2)}x^2 + \frac{1}{2}\frac{M_x-\nu M_z}{D(1-\nu^2)}z^2. 
\label{6-EqS7}
\end{equation}
\vspace{2pt}
%%%%%%%%%%%%%%%%%%%%%%%%%%%%%%%%%%%%%%%%%%%%%%%%%%
 
Note that this is equivalent to the deformed plate shape given by integrating the curvature equations and given by: %By substituting Equations (\ref{6-EqS5}) and (\ref{6-EqS6}) into Equation (\ref{6-EqS7}), the deformed plate equation can be described as a function of curvatures as follows:
%%%%%%%%%%%%%%%%%%%%%%%%%%%%%%%%%%%%%%%%%%%%%%%%%%
\vspace{8pt}
\begin{equation}
y = \frac{1}{2} (\kappa x^2 + \tau z^2). 
\label{6-EqS8}
\end{equation}
%%%%%%%%%%%%%%%%%%%%%%%%%%%%%%%%%%%%%%%%%%%%%%%%%%

\section{Kinematic Derivation: Synclastic Deformation (Both Curvatures Upward)} \label{AppendixB}

As shown in Figure \ref{6-Fig3} (a), the coordinate of points on the 1\textsuperscript{st} scale can be obtained according to coordinates $xyz$, which is located on the midpoint of 1\textsuperscript{st} scale's width $M_1$. The coordinate of these points are as follows:
%%%%%%%%%%%%%%%%%%%%%%%%%%%%%%%%%%%%%%%%%%%%%%%%%%
%\vspace{-5pt}
\begin{align}
& M_1 = (0,0,0) \hspace{12pt} , \hspace{5pt} N_1 = (-l \cos \theta, l \sin \theta , 0), \nonumber\\
& A_1 = (0,0,+b) \hspace{5pt} , \hspace{5pt} B_1 = (-l \cos \theta, l \sin \theta , +b), \label{6-EqS9}\\
& D_1 = (0,0,-b) \hspace{5pt} , \hspace{5pt} C_1 = (-l \cos \theta, l \sin \theta , -b). \nonumber
%\vspace{-15pt}
\end{align}
%%%%%%%%%%%%%%%%%%%%%%%%%%%%%%%%%%%%%%%%%%%%%%%%%%

Similar to the 1\textsuperscript{st} scale, the locations of points of the 2\textsuperscript{nd} scale before the deformation of the substrate can be obtained as follows:
%%%%%%%%%%%%%%%%%%%%%%%%%%%%%%%%%%%%%%%%%%%%%%%%%%
%\vspace{-5pt}
\begin{align}
& M_{2,0} = (d,0,-(b+a)) \hspace{10pt} , \hspace{5pt} N_{2,0} = (d-l \cos \theta, l \sin \theta , -(b+a)), \nonumber\\
& A_{2,0} = (d,0,-a) \hspace{40pt} , \hspace{5pt} B_{2,0} = (d-l \cos \theta, l \sin \theta , -a), \label{6-EqS10}\\
& D_{2,0} = (d,0,-(2b+a)) \hspace{5pt} , \hspace{5pt} C_{2,0} = (d-l \cos \theta, l \sin \theta , -(2b+a)). \nonumber
%\vspace{-5pt}
\end{align}
%%%%%%%%%%%%%%%%%%%%%%%%%%%%%%%%%%%%%%%%%%%%%%%%%%

According to \ref{AppendixA}, and Equation (\ref{6-EqS8}), the equation of deformed substrate under longitudinal bending curvature $\kappa=\frac{\psi}{d}$ and transverse bending curvature $\tau=\frac{\omega}{b+a}$ is as follows:
%%%%%%%%%%%%%%%%%%%%%%%%%%%%%%%%%%%%%%%%%%%%%
\begin{equation}
y = \frac{1}{2} \Big( \big(\frac{\psi}{d} \big)x^2 + \big(\frac{\omega}{b+a}\big)z^2 \Big).
%\vspace{-5pt}
\label{6-EqS11}
\end{equation}
%%%%%%%%%%%%%%%%%%%%%%%%%%%%%%%%%%%%%%%%%%%%%

Based on the Equation (\ref{6-EqS11}), the location of the midpoint of 2\textsuperscript{nd} scale's width $M_2$ after the deformation of substrate is as follows:
%%%%%%%%%%%%%%%%%%%%%%%%%%%%%%%%%%%%%%%%%%%%%
\begin{equation}
M_{2} = \Big(d,\frac{{\psi}{d}+{\omega}(b+a)}{2},-(b+a)\Big).
%\vspace{-5pt}
\label{6-EqS12}
\end{equation}
%%%%%%%%%%%%%%%%%%%%%%%%%%%%%%%%%%%%%%%%%%%%%

The longitudinal symmetry line of 2\textsuperscript{nd} scale's plate, $M_2N_2$, has an angle equal to $\theta$ with respect to the tangent line of the substrate at point $M_2$ in the direction of $-x$ axis. This mean the vector ${\bm {M}}_{\bm 2}{\bm N}_{\bm 2}$ with the length $l$ has an angle equal to $\theta-\psi$ with respect to the direction of $-x$ axis, which leads to: 
%%%%%%%%%%%%%%%%%%%%%%%%%%%%%%%%%%%%%%%%%%%%%
\begin{equation}
{\bm {M}}_{\bm 2}{\bm N}_{\bm 2}= \big( -l \cos (\theta-\psi), l \sin (\theta-\psi), l\sin (\theta-\psi)\sin\omega \big).
%\vspace{-5pt}
\label{6-EqS13}
\end{equation}
%%%%%%%%%%%%%%%%%%%%%%%%%%%%%%%%%%%%%%%%%%%%%

From Equation (\ref{6-EqS12}) and (\ref{6-EqS13}), the coordinate of point $N_2$ can be found as:
%%%%%%%%%%%%%%%%%%%%%%%%%%%%%%%%%%%%%%%%%%%%%
\begin{equation}
N_{2} = \Big(d-l \cos (\theta-\psi),\frac{{\psi}{d}+{\omega}(b+a)}{2}+l \sin (\theta-\psi),l\sin (\theta-\psi)\sin\omega-(b+a)\Big).
%\vspace{-5pt}
\label{6-EqS14}
\end{equation}
%%%%%%%%%%%%%%%%%%%%%%%%%%%%%%%%%%%%%%%%%%%%%

Because the edge $A_2D_2$ is tangent to the substrate surface at the middle point $M_2$, the angle of this edge with the direction of $z$ axis is equal to $\omega$. Also, line $A_2D_2$ is parallel to line $B_2C_2$. Based on this, the following vectors can be found as follows:
%%%%%%%%%%%%%%%%%%%%%%%%%%%%%%%%%%%%%%%%%%%%%
\begin{equation}
{\bm {A}}_{\bm 2}{\bm M}_{\bm 2}={\bm {M}}_{\bm 2}{\bm D}_{\bm 2}={\bm {B}}_{\bm 2}{\bm N}_{\bm 2}={\bm {N}}_{\bm 2}{\bm C}_{\bm 2}=(0,b \sin \omega,-b \cos \omega).
%\vspace{-5pt}
\label{6-EqS15}
\end{equation}
%%%%%%%%%%%%%%%%%%%%%%%%%%%%%%%%%%%%%%%%%%%%%

Based on the Equation (\ref{6-EqS12}) and (\ref{6-EqS15}), the coordinate of points $A_2$ and $D_2$ are as follows:
%%%%%%%%%%%%%%%%%%%%%%%%%%%%%%%%%%%%%%%%%%%%%%%%%%
%\vspace{-5pt}
\begin{align}
& A_2 = \Big(d,\frac{{\psi}{d}+{\omega}(b+a)}{2}-b \sin \omega,-(b+a)+b \cos \omega\Big), \nonumber\\
& D_2 = \Big(d,\frac{{\psi}{d}+{\omega}(b+a)}{2}+b \sin \omega,-(b+a)-b \cos \omega\Big). \label{6-EqS16}
%\vspace{-10pt}
\end{align}
%%%%%%%%%%%%%%%%%%%%%%%%%%%%%%%%%%%%%%%%%%%%%%%%%%

Also, according to the Equation (\ref{6-EqS14}) and (\ref{6-EqS15}), the locations of point $B_2$ and $C_2$ are as follows:
%%%%%%%%%%%%%%%%%%%%%%%%%%%%%%%%%%%%%%%%%%%%%%%%%%
%\vspace{-5pt}
\begin{align}
& B_2 = \Big(d-l \cos (\theta-\psi),\frac{{\psi}{d}+{\omega}(b+a)}{2}-b \sin \omega +l \sin (\theta-\psi),l\sin (\theta-\psi)\sin\omega-(b+a)+b \cos \omega\Big), \nonumber\\
& C_2 = \Big(d-l \cos (\theta-\psi),\frac{{\psi}{d}+{\omega}(b+a)}{2}+b \sin \omega +l \sin (\theta-\psi),l\sin (\theta-\psi)\sin\omega-(b+a)-b \cos \omega\Big). \label{6-EqS17}
%\vspace{-5pt}
\end{align}
%%%%%%%%%%%%%%%%%%%%%%%%%%%%%%%%%%%%%%%%%%%%%%%%%%

To find the equation of plate $A_1B_1C_1D_1$, the plate normal vector ${\bm {n}}_{\bm 1}=(x_{{\bm n}_{\bm 1}},y_{{\bm n}_{\bm 1}},z_{{\bm n}_{\bm 1}})$ is equal to:
%%%%%%%%%%%%%%%%%%%%%%%%%%%%%%%%%%%%%%%%%%%%%
\begin{equation}
{\bm {n}}_{\bm 1} = \frac{{\bm {M}}_{\bm 1}{\bm D}_{\bm 1} \times {\bm {M}}_{\bm 1}{\bm N}_{\bm 1}}{|{\bm {M}}_{\bm 1}{\bm D}_{\bm 1}||{\bm {M}}_{\bm 1}{\bm N}_{\bm 1}|} = ( \sin \theta , \cos \theta , 0 )
%\vspace{-5pt}
\label{6-EqS18}
\end{equation}
%%%%%%%%%%%%%%%%%%%%%%%%%%%%%%%%%%%%%%%%%%%%%

The general form of plate equation with normal vector ${\bm {n}}_{\bm 1}$ and point $M_1$ located in the plate is described as $x_{{\bm n}_{\bm 1}}\big( x-x_{M_1}\big)+y_{{\bm n}_{\bm 1}}\big( y-y_{M_1}\big)+z_{{\bm n}_{\bm 1}}\big( z-z_{M_1}\big)=0$. By using the plate normal vector from Equation (\ref{6-EqS18}) and coordinates of point $M_1$, equation of plate $A_1B_1C_1D_1$ is as follows:
%%%%%%%%%%%%%%%%%%%%%%%%%%%%%%%%%%%%%%%%%%%%%
\begin{equation}
x\sin \theta + y \cos \theta = 0 \hspace{5pt}
%\vspace{-5pt}
\label{6-EqS19}
\end{equation}
%%%%%%%%%%%%%%%%%%%%%%%%%%%%%%%%%%%%%%%%%%%%%

To find the contact between the corner of 2\textsuperscript{nd} scale $B_2$ and surface of 1\textsuperscript{st} scale, the location of point $B_2$ shown in Equation (\ref{6-EqS17}), must satisfy the $A_1B_1C_1D_1$ plate equation, which yields:
%%%%%%%%%%%%%%%%%%%%%%%%%%%%%%%%%%%%%%%%%%%%%
\begin{equation}
\Big(\hspace{-2pt}d-l \cos (\theta-\psi) \Big) \sin \theta + \Big( \frac{{\psi}{d} + {\omega}(b+a)}{2}-b \sin \omega +l \sin (\theta-\psi) \Big)\cos \theta= 0 \hspace{5pt}
%\vspace{-5pt}
\label{6-EqS20}
\end{equation}
%%%%%%%%%%%%%%%%%%%%%%%%%%%%%%%%%%%%%%%%%%%%%

By nondimensionalization of the geometrical parameters with respect to the longitudinal spacing between the scale $d$, we define dimensionless geometric parameters including $\eta = l/d$, $\beta = b/d$, and $\delta = a/d$. Using these dimensionless parameters, the Equation (\ref{6-EqS20}) is rewritten as:
%%%%%%%%%%%%%%%%%%%%%%%%%%%%%%%%%%%%%%%%%%%%%
\begin{equation}
\eta \sin \psi - \sin \theta - \cos \theta \Big(\frac{\psi}{2}+\big(\frac{\beta + \delta}{2}\big) \omega - \beta \sin \omega \Big) = 0.
\label{6-EqS21}
\end{equation}
%%%%%%%%%%%%%%%%%%%%%%%%%%%%%%%%%%%%%%%%%%%%%

Now, to analyze the longitudinal locking response of the plate, we took derivative of Equation (\ref{6-EqS21}) which is:

%%%%%%%%%%%%%%%%%%%%%%%%%%%%%%%%%%%%%%%%%%%%%
\begin{equation}
 \frac{\partial\psi}{\partial\theta}= \frac{\sin\theta(\frac{\psi}{2}+\frac{\beta+\delta}{2}\omega-\beta \sin\omega)-\cos\theta}{\frac{1}{2}\cos\theta-\eta\cos\psi}.
\label{6-EqS22}
\end{equation}
%%%%%%%%%%%%%%%%%%%%%%%%%%%%%%%%%%%%%%%%%%%%%

To find the locking region of the plate, we made $\frac{\partial\psi}{\partial\theta}$ = 0, and thus, 
%%%%%%%%%%%%%%%%%%%%%%%%%%%%%%%%%%%%%%%%%%%%%
\begin{equation}
\sin\theta(\frac{\psi}{2}+\frac{\beta+\delta}{2}\omega-\beta \sin\omega)-\cos\theta = 0.
\label{6-EqS23}
\end{equation}
%%%%%%%%%%%%%%%%%%%%%%%%%%%%%%%%%%%%%%%%%%%%%

The plate structures will be transversely locked when two scales touch each other in the transverse direction. As we see in Figure \ref{6-Fig3} (a), transverse locking will happen when point $B_2$ is on $x$-axis. Thus, by making $z$-coordinate of $B_2$ equal to zero:
%%%%%%%%%%%%%%%%%%%%%%%%%%%%%%%%%%%%%%%%%%%%%%%%%%
\begin{equation}
l\sin (\theta-\psi)\sin\omega-(b+a)+b \cos \omega = 0
\label{6-EqS24}
\end{equation}
%%%%%%%%%%%%%%%%%%%%%%%%%%%%%%%%%%%%%%%%%%%%%%%%%%

and then non-dimensionalizing it, we will get the following expression of $\theta$:

%%%%%%%%%%%%%%%%%%%%%%%%%%%%%%%%%%%%%%%%%%%%%%%%%%
\begin{equation}
\theta = \sin^{-1}\left(\frac{\delta + (1 - \cos\omega)\beta}{\eta\sin\omega}\right) + \psi
\label{6-EqS25}
\end{equation}
%%%%%%%%%%%%%%%%%%%%%%%%%%%%%%%%%%%%%%%%%%%%%%%%%%
Now, to obtain locking curvature in the transverse direction, we substitute the expression of $\theta$ from Equation (\ref{6-EqS25}) into Equation (\ref{6-EqS21}):

%%%%%%%%%%%%%%%%%%%%%%%%%%%%%%%%%%%%%%%%%%%%%%%%%%%%%%%%%%%%%%%%%
\begin{equation}
\begin{aligned}
\eta\sin\psi - \sin \left\{ \sin^{-1} \left( \frac{\delta + (1 - \cos\omega)\beta}{\eta\sin\omega}\right) + \psi  \right\} - \cos \left\{ \sin^{-1} \left( \frac{\delta + (1 - \cos\omega)\beta}{\eta\sin\omega}\right) + \psi  \right\} \\
\left\{ \frac{\psi}{2} + \left( \frac{\beta + \delta}{2} \right)\omega - \beta\sin\omega \right\} = 0 \label{6-EqS26}
\end{aligned}
\end{equation}
%%%%%%%%%%%%%%%%%%%%%%%%%%%%%%%%%%%%%%%%%%%%%%%%%%%%%%%%%%%%%%%%

\section{Kinematic Derivation: Anti-clastic deformation (Longitudinal Curvature Upward and Transverse Curvature Downward)} \label{AppendixC}

In this section, we use the locations of points $C_1$, $D_1$, $B_2$, and $C_2$, which is provided in Equation (\ref{6-EqS9}) and Equation (\ref{6-EqS17}) to find the equation of lines $D_1C_1$ and $B_2C_2$. For this purpose, the directional vector of line $D_1C_1$ is called ${\bm {m}}_{\bm 1}=(x_{{\bm m}_{\bm 1}},y_{{\bm m}_{\bm 1}},z_{{\bm m}_{\bm 1}})$, and the directional vector of line $B_2C_2$ is called ${\bm {m}}_{\bm 2}=(x_{{\bm m}_{\bm 2}},y_{{\bm m}_{\bm 2}},z_{{\bm m}_{\bm 2}})$. These vectors are as follows:
%%%%%%%%%%%%%%%%%%%%%%%%%%%%%%%%%%%%%%%%%%%%%
\begin{align}
& {\bm {m}}_{\bm 1} = \frac{{\bm {D}}_{\bm 1}{\bm C}_{\bm 1}}{|{\bm {D}}_{\bm 1}{\bm C}_{\bm 1}|} = ( -\cos \theta , \sin \theta , 0 ), \label{6-EqS27}\\
& {\bm {m}}_{\bm 2} = \frac{{\bm {B}}_{\bm 2}{\bm C}_{\bm 2}}{|{\bm {B}}_{\bm 2}{\bm C}_{\bm 2}|} = ( 0 , \sin \omega , -\cos \omega ). \label{6-EqS28}
%\vspace{-5pt}
\end{align}
%%%%%%%%%%%%%%%%%%%%%%%%%%%%%%%%%%%%%%%%%%%%%

By using the directional vectors and one point located on each line, the general 3D-equation of the lines $D_1C_1$ and $B_2C_2$ are as follows:

%%%%%%%%%%%%%%%%%%%%%%%%%%%%%%%%%%%%%%%%%%%%%
\begin{align}
& \frac{x-x_{D_1}}{x_{{\bm m}_{\bm 1}}}=\frac{y-y_{D_1}}{y_{{\bm m}_{\bm 1}}}=\frac{z-z_{D_1}}{z_{{\bm m}_{\bm 1}}}=p, \label{6-EqS29}\\
& \frac{x-x_{C_2}}{x_{{\bm m}_{\bm 2}}}=\frac{y-y_{C_2}}{y_{{\bm m}_{\bm 2}}}=\frac{z-z_{C_2}}{z_{{\bm m}_{\bm 2}}}=q. \label{6-EqS30}
\end{align}
%\vspace{-25pt}
%%%%%%%%%%%%%%%%%%%%%%%%%%%%%%%%%%%%%%%%%%%%%

The parametric form of the equation of line $D_1C_1$ as follows, where $p$ can vary from $0$ to $l$:
%%%%%%%%%%%%%%%%%%%%%%%%%%%%%%%%%%%%%%%%%%%%%
%\vspace{-10pt}
\begin{align}
& x(p) = -p\cos \theta, \nonumber \\
& y(p) = p\sin \theta, \label{6-EqS31}\\
& z(p) = -b. \nonumber
\end{align}
%%%%%%%%%%%%%%%%%%%%%%%%%%%%%%%%%%%%%%%%%%%%%

The parametric form of the equation of line $B_2C_2$ as follows, where $q$ can vary from $-2b$ to $0$:
%%%%%%%%%%%%%%%%%%%%%%%%%%%%%%%%%%%%%%%%%%%%%
\begin{align}
& x(q) = d-l \cos (\theta-\psi), \nonumber \\
& y(q) = q \sin \omega + \frac{{\psi}{d}+{\omega}(b+a)}{2}+b \sin \omega +l \sin (\theta-\psi), \label{6-EqS32} \\ 
& z(q) = -q \cos \omega -(b+a)-b \cos \omega. \nonumber
\end{align}
%%%%%%%%%%%%%%%%%%%%%%%%%%%%%%%%%%%%%%%%%%%%%

To find a contact point between these two lines, Equations (\ref{6-EqS31}) and (\ref{6-EqS32}) must be identical at $x$, $y$, and $z$ coordinate simultaneously, which means $x(p)=x(q)$, $y(p)=y(q)$, and $z(p)=z(q)$. From $x(p)=x(q)$, we find a relationship for $p$, and from $z(p)=z(q)$, we obtain a relationship for $q$. By substituting the obtained relationships for $p$ and $q$ into $y(p)=y(q)$, we find the following equation:
%%%%%%%%%%%%%%%%%%%%%%%%%%%%%%%%%%%%%%%%%%%%%
%\vspace{-10pt}
\begin{equation}
\Big(\frac{l \cos(\theta-\psi) - d}{\cos \theta} \Big)\sin \theta = \Big(\frac{a+b\cos \omega}{-\cos \omega} \Big) \sin \omega + \frac{{\psi}{d}+{\omega}(b+a)}{2}+b \sin \omega +l \sin (\theta-\psi). \hspace{5pt}
%\vspace{-5pt}
\label{6-EqS33}
\end{equation}
%%%%%%%%%%%%%%%%%%%%%%%%%%%%%%%%%%%%%%%%%%%%%

By nondimensionalization the geometrical parameters with respect to $d$, the Equation (\ref{6-EqS33}) can be rewritten as:
%%%%%%%%%%%%%%%%%%%%%%%%%%%%%%%%%%%%%%%%%%%%%
\begin{equation}
\eta \sin \psi - \sin \theta - \cos \theta \Big(\frac{\psi}{2}+\big(\frac{\beta + \delta}{2}\big) \omega - \delta \tan \omega \Big) = 0. 
\label{6-EqS34}
\end{equation}
%%%%%%%%%%%%%%%%%%%%%%%%%%%%%%%%%%%%%%%%%%%%%

This relationship is slightly different than Equation (\ref{6-EqS21}). Similar to Equation (\ref{6-EqS21}), to analyze the locking response of the plate we took derivative of Equation (\ref{6-EqS34}) which is:
%%%%%%%%%%%%%%%%%%%%%%%%%%%%%%%%%%%%%%%%%%%%%
\begin{equation}
 \frac{\partial\psi}{\partial\theta}= \frac{\sin\theta(\frac{\psi}{2}+\frac{\beta+\delta}{2}\omega-\delta \tan\omega)-\cos\theta}{-\eta\cos\theta+\frac{1}{2}\cos\theta}.
\label{6-EqS35}
\end{equation}
%%%%%%%%%%%%%%%%%%%%%%%%%%%%%%%%%%%%%%%%%%%%%
To find the locking region of the plate, we made, $\frac{\partial\psi}{\partial\theta}$ = 0, and thus, 
%%%%%%%%%%%%%%%%%%%%%%%%%%%%%%%%%%%%%%%%%%%%%
\begin{equation}
\sin\theta(\frac{\psi}{2}+\frac{\beta+\delta}{2}\omega-\delta \tan\omega)-\cos\theta = 0.
\label{6-EqS36}
\end{equation}
%%%%%%%%%%%%%%%%%%%%%%%%%%%%%%%%%%%%%%%%%%%%%
%%%%%%%%%%%%%%%%%%%%%%%%%%%%%%%%%%%%%%%%%%%%%%%%%%%

\section{Derivation of inclusion correction factors $C_{f,x}$ and $C_{f,z}$} \label{AppendixD}

The moment-curvature relationship of a plain Kirchoff plate in linear elastic framework can be written as \cite{reddy2006theory}:

%%%%%%%%%%%%%%%%%%%%%%%%%%%%%
\begin{equation}
\begin{Bmatrix} M_z \\ M_x \end{Bmatrix} = D\begin{bmatrix} 1 & \nu \\ \nu & 1\end{bmatrix}\begin{Bmatrix} \kappa \\ \tau \end{Bmatrix}
\end{equation}
\label{6-Eq7a}
%%%%%%%%%%%%%%%%%%%%%%%%%%%%%%%%%%%%%
Again, $D$ is the bending rigidity of the plain plate, and $\nu$ is the Poisson's ratio. The scales on the substrate have an embedded part that increases the stiffness of the substrate even before engagement is achieved. This is the so-called inclusion effect, which in this case of plate changes the isotropic substrate into a composite structure. The increase in bending rigidity due to these inclusions can be modeled empirically by assuming two inclusion correction factors $C_{f,x}$, and $C_{f,z}$, similar to previous works on 1D substrates \cite{ebrahimi2019tailorable}. The scale-embedded plate can be modeled as a short-fiber orthotropic composite plate and thus the correction factors then lead to a new modified moment-curvature relationship:

%The presence of rigid scales on substrate is modeled as short fiber orthotropic composite material. Thus, from the basic knowledge of composite we assumed that the diagonal term of stiffeness matrix will be changes. Due to the presence of rigid scales in the soft substrate, the rigidity of the substrate will be increased and this effect is addressed by two inclusion factors $C_{f,x}$ and $C_{f,z}$ in $x$ and $z$-directions, respectively. From previous 1D and FEM results analysis we found that the unit term in stiffness matrix will be replaced with $C_{f,x}$ and $C_{f,z}$. Thus, considering the inclusion correction factor the stiffness matrix can be written as:}

%%%%%%%%%%%%%%%%%%%%%%%%%%%%%
\begin{equation}
\begin{Bmatrix} M_z \\ M_x \end{Bmatrix} = D\begin{bmatrix} C_{f,x} & \nu \\ \nu & C_{f,z}\end{bmatrix}\begin{Bmatrix} \kappa \\ \tau \end{Bmatrix}
\end{equation}
\label{6-Eq7b}
%%%%%%%%%%%%%%%%%%%%%%%%%%%%%%%%%%%%%

Now, from classical plate theory, the strain energy stored in the plate due to bi-directional bending is \cite{kelly2013solid}:
% %%%%%%%%%%%%%%%%%%%%%%%%%%%%%%%%%%%%%%%%%%%%

 \begin{equation}
 \Delta U = \frac{1}{2}(M_z \kappa + M_x\tau) L_{sub} W_{sub} \label{6-Eq6a}
 \end{equation}
 %%%%%%%%%%%%%%%%%%%%%%%%%%%%%%%%%%%%%%%%%%%%%
 
Substituting the expressions of $M_x$ and $M_z$ in Equation (\ref{6-Eq6a})
 %%%%%%%%%%%%%%%%%%%%%%%%%%%%%%%%%%%%%%%%%%%%%%%%%%%

\begin{equation}
\Delta U = \frac{D}{2} (C_{f,x} \kappa^2+ 2\nu \tau\kappa +C_{f,z} \tau^2) L_{sub} W_{sub}. \label{6-Eq6b}
\end{equation}
 
% %%%%%%%%%%%%%%%%%%%%%%%%%%%%%%%%%%%%%%%%%%%%

% To obtain relationships for inclusion correction factors $C_{f,x}$ and $C_{f,z}$, for two in-plane directions, extensive finite-element simulations are performed. $C_{f,x}$ and $C_{f,z}$ are found to be depended on the size, shape, and volume fraction of the inclusion, and two different functions based on three embedded parameters ($L$, $D$, and $b$) are considered as: $C_{f,x} = C_{f,x1} + C_{f,x2} \big({\zeta \beta \gamma}\big) h(\theta_{o})$  and $C_{f,z} = C_{f,z1} + C_{f,z2} ln\big(\frac{\zeta \beta \gamma}{\delta}\big) h(\theta_{o})$.  

\begin{figure} [htbp]
\begin{center}
\includegraphics[scale = 0.75]{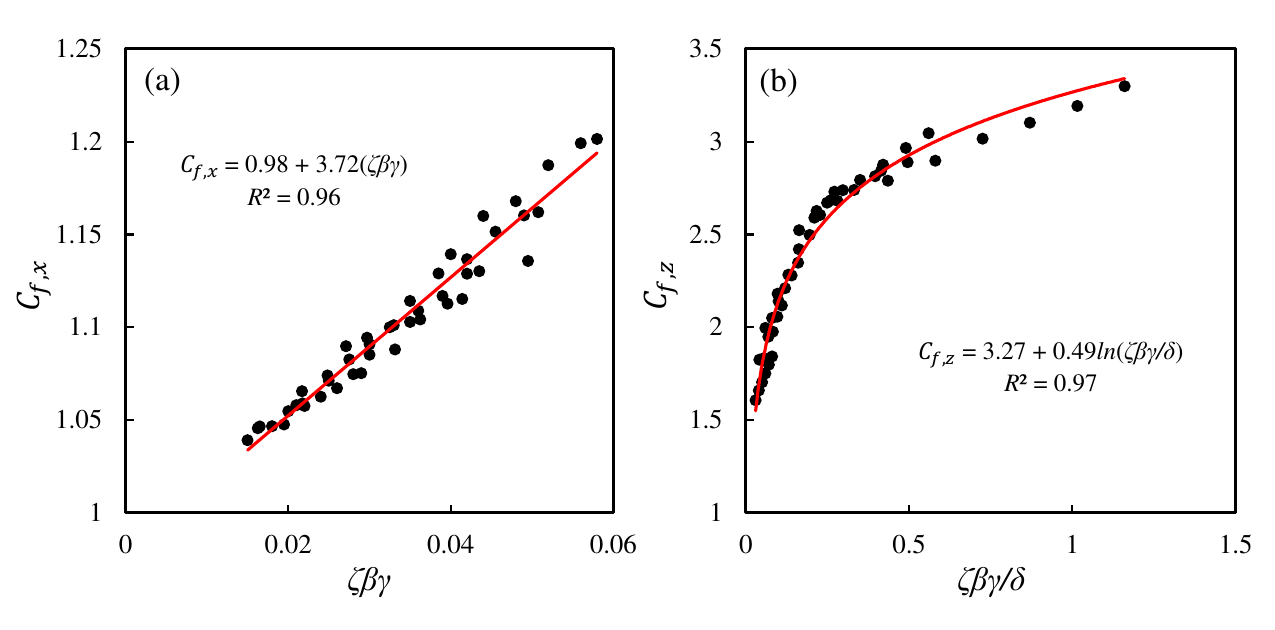}
\end{center}
%\vspace{-20pt}
\caption
{\label{6-FigD17} Dimensionless inclusion correction factor (a) $C_{f,x}$, (b) $C_{f,z}$ as a function of dimensionless geometrical  variable group.}
\end{figure}

To obtain the expressions of $C_{f,x}$ and $C_{f,z}$, a lot of finite-element simulations are carried out varying all three embedded dimensions, $L, t_s,$ and $b$. The exposed length of the scale is considered to be 0 (so, $l$ = 0, and therefore $\eta$ = 0). If there is no scale embedded in the plate, $C_{f,x}$ and $C_{f,z}$ are equal to 1. But for embedded scaled plate both $C_{f,x}$ and $C_{f,z}$ are expected to be $\ge$ 1. Thus, two different equations for $C_{f,x}$ and $C_{f,z}$ as a function of four dimensionless parameters, $\zeta, \beta, \gamma$, and $\delta$ are considered as follows:

%%%%%%%%%%%%%%%%%%%%%%%%%%%%%%%%%%%%%%%%%%%%%%%%%
\begin{equation}
C_{f,x} = C_{f,x1} + C_{f,x2} \big({\zeta \beta \gamma}\big) h(\theta_{o}),
\end{equation} %\vspace{-22pt}

\begin{equation}
C_{f,z} = C_{f,z1} + C_{f,z2} ln\big(\frac{\zeta \beta \gamma}{\delta}\big) h(\theta_{o}).
\end{equation}
%%%%%%%%%%%%%%%%%%%%%%%%%%%%%%%%%%%%%%%%%%%%%%%%%%
Here, two dimensionless parameters $\zeta = L/d$ and $\gamma = \sqrt{t_s/d}$. $C_{f,x1}$, $C_{f,x2}$, $C_{f,z1}$, and $C_{f,z2}$ are arbitrary constants which is determined using finite-element simulations as shown in Figure (\ref{6-FigD17}). In Figure (\ref{6-FigD17}), $C_{f,x}$ and $C_{f,z}$ is plotted from FE data by varying $\zeta, \beta, \gamma$, and $\delta$, and fitting the best equations. From Figure (\ref{6-FigD17}), the constants value can be calculated as $C_{f,x1} = 0.98$, $C_{f,x2}= 3.72$, $C_{f,z1}= 3.27$, and $C_{f,z2}= 0.49$. And in both $C_{f,x}$ and $C_{f,z}$, dimensionless angular function $h(\theta_{o})$ is found to be $\approx$ 1 which shows that initial scale inclination angle $\theta_o$ doesn't have any significant effect in inclusion correction factors.  These two equations are then implemented in analytical expressions developed in Equations (\ref{6-Eq5}) and (\ref{6-Eq6}) to calculate the moment-curvature relation of the scale-covered plate.   
%%%%%%%%%%%%%%%%%%%%%%%%%%%%%%%%%%%%%%%%%%%%%%%%%%
%\vspace{-10pt}
%\vspace{-10pt}
%%%%%%%%%%%%%%%%%%%%%%%%%%%%%%%%%%%%%%%%%%%%%%%%%%

 
 \bibliographystyle{elsarticle-num} 
 \bibliography{cas-refs}





% \end{thebibliography}
\end{document}